\documentclass[twocolumn,showpacs,preprintnumbers,amsmath,amssymb]{revtex4}

\usepackage{graphicx}
\usepackage{dcolumn}
\usepackage{bm}

\newcommand{\al}{\alpha}
\newcommand{\be}{\beta}
\newcommand{\ga}{\gamma}

\newcommand{\de}{\delta}
\newcommand{\De}{\Delta}

\newcommand{\eps}{\epsilon}

\newcommand{\Om}{\Omega}

\newcommand{\p}{\partial}
 
\newcommand{\txt}{\textstyle}
\newcommand{\dsp}{\displaystyle}

\newcommand\eqn[1]{(\ref{#1})}      
\newcommand\Eqn[1]{Eq.~(\ref{#1})}  
\newcommand{\e}{{\rm e}}   
\newcommand{\beq}{\begin{equation}}
\newcommand{\eeq}{\end{equation}}
\newcommand{\ba}{\begin{array}}
\newcommand{\ea}{\end{array}}
\newcommand{\bea}{\begin{eqnarray}}
\newcommand{\eea}{\end{eqnarray}}
\newcommand{\bi}{\begin{itemize}}  
\newcommand{\ei}{\end{itemize}}
\newcommand{\ben}{\begin{enumerate}} 
\newcommand{\een}{\end{enumerate}}

\newcommand{\half} {{\txt \frac{1}{2}}}
\newcommand{\third}{{\txt \frac{1}{3}}}

\newcommand{\twothirds}{{\txt \frac{2}{3}}}

\newcommand\hide[1]{}

\newcommand{\Det}{\mbox{Det}}

\newcommand{\psibar}{{\bar\psi}}

\newcommand{\feyn}[1]{
  \setbox0=\hbox{\ensuremath{#1}}
  \hbox to\wd0{\hbox to0pt{\hbox to\wd0{\hss/\hss}\hss}\box0}}

\newcommand{\keV}{{\rm keV}} 
\newcommand{\MeV}{{\rm MeV}}

\newcommand{\diag}{{\rm diag}}

\newcommand{\Qtilde}{{\tilde Q}}

\newcommand{\Drugdbs}{D_{\mbox{\scriptsize$\!\!\!\ba{c}ru\\[-0.5ex]gd\\[-0.3ex]bs\ea $}}}
\newcommand{\Drdgu}{D_{\mbox{\scriptsize$\!\!\ba{c}rd\\[-0.5ex]gu\ea $}}}
\newcommand{\Drsbu}{D_{\mbox{\scriptsize$\!\!\ba{c}rs\\[-0.5ex]bu\ea $}}}
\newcommand{\Dgsbd}{D_{\mbox{\scriptsize$\!\!\ba{c}gs\\[-0.5ex]bd\ea $}}}

\begin{document}

\preprint{MIT-CTP-3503}

\title{Evaluating the Gapless Color-Flavor Locked Phase}

\author{Mark Alford}
\affiliation{Physics Department, Washington University,
St.~Louis, MO~63130, USA}
\author{Chris Kouvaris}\author{Krishna Rajagopal}
\affiliation{Center for Theoretical Physics, Massachusetts 
Institute of Technology, Cambridge, MA 02139, USA}

\date{June 11, 2004}

\begin{abstract}
In neutral cold quark matter that is sufficiently dense that the
strange quark mass $M_s$ is unimportant, all nine quarks (three
colors; three flavors) pair in a color-flavor locked (CFL) pattern,
and all fermionic quasiparticles have a gap. We recently argued that
the next phase down in density (as a function of decreasing quark
chemical potential $\mu$ or increasing strange quark mass $M_s$) is
the new ``gapless CFL'' (``gCFL'') phase in which only 
seven quasiparticles have
a gap, while there are gapless quasiparticles described by
two dispersion relations at three momenta.
There is a continuous quantum phase transition from CFL to
gCFL quark matter at $M_s^2/\mu\approx 2\Delta$,
with $\Delta$ the gap parameter.  Gapless CFL, like CFL, leaves
unbroken a linear combination $\tilde Q$ of electric and color
charges, but it is a $\tilde Q$-conductor with 
gapless $\tilde Q$-charged quasiparticles and a nonzero electron density.
In this paper, we evaluate the gapless CFL phase, in several senses.
We present the details underlying our earlier work which showed how
this phase arises.
We display all nine quasiparticle dispersion relations in full detail.
Using a general pairing ansatz that only neglects effects that are
known to be small, we perform a comparison of the free energies of the
gCFL, CFL, 2SC, gapless 2SC, and 2SCus phases.  We conclude that
as density drops, making the CFL phase less favored, the gCFL phase is
the next spatially uniform quark matter phase to occur.  A mixed phase
made of colored components would have lower free energy if color were
a global symmetry, but in QCD such a mixed phase is penalized
severely.

\end{abstract}


\maketitle

\section{Introduction}

Because QCD is asymptotically free,
we expect that matter at sufficiently high densities and/or
temperatures
will consist of almost-free quarks and gluons. However, over the
last few years it has become clear that 
there is a rich and varied landscape of phases
lying between these asymptotic
regimes and the familiar hadronic
phase at low temperature and density.
In the region where the temperature is low and the density is
high enough that hadrons are crushed into quark matter, 
there is a whole family of ``color superconducting'' phases~\cite{reviews}.
The essence of color superconductivity is quark pairing, driven by
the BCS mechanism, which operates when there exists
an attractive interaction between fermions at a Fermi surface.
The QCD quark-quark interaction is strong, and is attractive
in many channels, so we expect cold dense quark matter to {\em generically}
exhibit color superconductivity.
Moreover, quarks, unlike electrons, have color and flavor as well as spin
degrees of freedom, so many different patterns of pairing are possible.
This leads us to expect a rich phase structure
in matter beyond nuclear density. 

Color superconducting quark matter may well occur naturally in the
universe, in the cold dense cores of compact (``neutron'') stars,
where densities are above nuclear density, and temperatures are of the
order of tens of {\rm keV}.  
In future low-energy heavy ion colliders, such as 
the Compressed
Baryonic Matter Experiment
at the future accelerator facility at GSI Darmstadt~\cite{GSI}, 
it could conceivably be possible to
create color superconducting quark matter 
(or perhaps hot dense matter that is in the quark-gluon
plasma phase but which exhibits fluctuations that are 
precursors of color 
superconductivity~\cite{Kitazawa:2001ft}).

It is by now well-established
that at asymptotic densities, where the up, down and strange
quarks can be treated on an equal footing and the potentially disruptive
strange quark mass can can be neglected, quark matter is in
the  color-flavor locked (CFL) phase,
in which quarks of 
all three colors and all three flavors form Cooper pairs~\cite{Alford:1998mk}.
However,  
just as RHIC is teaching us about the 
properties of the hot but
far from asymptotically hot 
quark-gluon plasma~\cite{Gyulassy:2004zy},
we should expect that if neutron star
cores are made of color superconducting quark matter, they may not
reach the densities at which CFL predominates.
In this paper, as in  
Ref.~\cite{Alford:2003fq}, we ask what form of 
color superconducting quark matter is the ``next phase down
in density''.  That is, we imagine beginning in the CFL phase at asymptotic
density, reducing the density, and assume that CFL pairing is 
disrupted by the heaviness of the strange quark before
color superconducting quark matter is superseded by the hadronic
phase.  Upon making this assumption, we ask what form the
disruption takes and what are the properties of the 
resulting phase of dense, but not asymptotically dense, matter.

To describe quark matter as may exist in the cores of 
compact stars,
we consider quark chemical potentials $\mu$ of order $500$ MeV at most.
The strange quark mass $M_s$ must then be included: it is
expected to be density dependent, lying between
the current mass $\sim 100$ MeV and the vacuum constituent quark
mass $\sim 500$~MeV.  
In bulk matter, as is relevant for 
compact stars where we are interested in
kilometer-scale volumes, we must furthermore require
electromagnetic and color neutrality \cite{BaymIida,Alford:2002kj} (possibly
via mixing of oppositely-charged phases) and allow for equilibration
under the weak interaction.  All
these factors work to pull apart the Fermi momenta of
the different quark species, imposing an energy cost on the cross-species
pairing that characterizes color-flavor locking. 
At the highest densities we expect CFL pairing, but as the
density decreases the combination of nonzero $M_s$ and the constraints
of neutrality put greater and greater stress on cross-species pairing,
and we expect transitions to other pairing patterns.

In this paper we study the first of these transitions, and
work exclusively at zero temperature, 
which is a reasonable
approximation in the interior of a neutron star
that is more than a few seconds old. 
(Nonzero
temperature adds interesting new facets to the analysis~\cite{Ruester:2004eg},
that we shall further analyze elsewhere.)
We argue that the CFL phase will first give way, via a
continuous phase transition, to a new phase with gapless fermions
that we call gapless CFL (gCFL).  The transition occurs when
$M_s^2/\mu \simeq 2\Delta$, where $\Delta$ is
the pairing gap parameter. The gCFL phase
has gapless modes and nonzero
electron density. Although it has the same symmetries 
as the CFL phase, gapless CFL
matter is a conductor whereas CFL quark matter is a
dielectric insulator.

\subsection{Summary of the CFL phase}

To set the stage for our analysis, we briefly summarize the properties
of the CFL phase~\cite{Alford:1998mk}. 
If we set all three quark masses to zero,
the diquark condensate in the CFL phase spontaneously breaks the
full symmetry group of QCD, 
\beq
\ba{r@{}l}
 {[SU(3)_{\rm color}]} &\times 
  \underbrace{SU(3)_L \times SU(3)_R}_{\displaystyle\supset [U(1)_Q]}
 \times U(1)_B \\[5ex]
 &\to\quad 
  \underbrace{SU(3)_{C+L+R}}_{\displaystyle\supset [U(1)_{{\tilde Q} }]} 
  \times \mathbb{Z}_2
\ea
\eeq
where $SU(3)_{\rm color}$ and
electromagnetism $U(1)_Q$ are gauged, and the unbroken
$SU(3)_{c+L+R}$ subgroup consists of 
flavor rotations of the left and right quarks 
with equal and opposite color rotations, and contains an unbroken
gauged ``rotated electromagnetism'' 
$U(1)_{\rm \tilde Q}$~\cite{Alford:1998mk,Alford:1999pb}. 
The CFL phase has the largest possible unbroken symmetry
consistent with diquark condensation, achieved by having
all nine quarks participate equally in the
pairing,  and this gives the maximal pairing free energy
benefit. Not surprisingly, {\em ab initio}
calculations valid at asymptotic densities confirm that the
CFL phase is the ground state of QCD in the high density
limit~\cite{Evans:1999at,reviews}. 

In the limit of three massless quarks described above
there are 17 broken symmetry generators in the CFL phase, 8 of which become
longitudinal components of massive gauge bosons and 9 of which
remain as Goldstone bosons.  However,
in the real world there are two light quark flavors, the up 
($u$) and down ($d$), with
masses $\lesssim 10~{\rm MeV}$, and a medium-weight flavor, the strange 
($s$) quark, with mass $\gtrsim 100~{\rm MeV}$.
The strange quark therefore plays a crucial role in the phases of QCD.
In the presence of quark masses, the eight Goldstone bosons
coming from the breaking of chiral symmetries acquire
masses~\cite{Alford:1998mk,Son:1999cm,reviews}, 
and furthermore the CFL condensate
may rotate within the manifold describing these mesons~\cite{Bedaque:2001je}.
In analyzing the response of the CFL phase to the strange
quark mass, we shall be concerned with the dispersion relations
describing its fermionic quasiparticles, as they signal an
instability corresponding to the disruption of pairing itself.
In this analysis, we shall neglect flavor rotations of the
CFL condensate, as the direct effects of such meson condensates
on the stability or instability with respect to pair breaking
is minimal. (Meson condensates would play an important indirect
role if they were charged, but the favored meson condensation
channels are neutral assuming that the neutrino density is
negligible~\cite{Bedaque:2001je}.)  The ninth Goldstone boson,
that corresponding
to the spontaneous breaking of $U(1)_B$ and hence to 
superfluidity, remains massless even once quark masses
are taken into account 
and therefore plays a crucial role in many low
energy properties of the CFL phase, for example in its
viscosity \cite{viscosity},
specific heat, neutrino opacity, and
neutrino emissivity at low temperatures~\cite{neutrinos}.

\subsection{(Gapless) CFL Pairing Ansatz}
\label{sec:ansatz}

To study the response of the CFL phase to a non-negligible strange
quark mass, we use the pairing ansatz~\cite{Alford:1998mk}
\begin{equation}
\langle \psi^\alpha_a C\gamma_5 \psi^\beta_b \rangle \sim 
\Delta_1 \eps^{\alpha\beta 1}\eps_{ab1} \!+\! 
\Delta_2 \eps^{\alpha\beta 2}\eps_{ab2} \!+\! 
\Delta_3 \eps^{\alpha\beta 3}\eps_{ab3}
\label{condensate}
\end{equation} 
Here $\psi^\alpha_a$ is a quark of color $\alpha=(r,g,b)$ 
and flavor $a=(u,d,s)$;
the condensate is a Lorentz scalar, antisymmetric in Dirac indices,
antisymmetric in color 
(the channel with the strongest
attraction between quarks), and consequently
antisymmetric in flavor. 
The gap parameters
$\De_1$, $\De_2$ and $\De_3$ describe down-strange,
up-strange and up-down Cooper pairs, respectively.
They describe a $9\times 9$ matrix in color-flavor
space that, in the basis $(ru, gd, bs, rd, gu, rs, bu, gs, bd)$,
takes the form
\beq
\label{blocks}
\bm{\Delta}= 
\newcommand{\mDe}{\makebox[1.2em][r]{$-\De$}} 
\left(
\begin{array}{ccccccccc}
 0 & \Delta_3 & \Delta_2 & 0 & 0 & 0 & 0 & 0 & 0 \\
 \Delta_3 & 0 & \Delta_1 & 0 & 0 & 0 & 0 & 0 & 0 \\
\Delta_2 & \Delta_1 & 0  & 0 & 0 & 0 & 0 & 0 & 0 \\
  0 & 0 & 0 & 0 & \mDe_3 & 0 & 0 & 0 & 0 \\
  0 & 0 & 0 & \mDe_3 & 0 & 0 & 0 & 0 & 0 \\
  0 & 0 & 0 & 0 & 0 & 0 & \mDe_2 & 0 & 0 \\
  0 & 0 & 0 & 0 & 0 & \mDe_2 & 0 & 0 & 0 \\
  0 & 0 & 0 & 0 & 0 & 0 & 0 & 0 & \mDe_1 \\
  0 & 0 & 0 & 0 & 0 & 0 & 0 & \mDe_1 & 0 \\
\end{array}
\right)
\eeq
We see that $(rd,gu)$,
$(bu,rs)$ and $(gs,bd)$
quarks pair with gap parameters $\De_1$, $\De_2$ and $\De_3$
respectively, while the $(ru,gd,bs)$ quarks pair among each other
involving all the $\De$'s. The most important physics that we
are leaving out by making this ansatz is pairing in which
the Cooper pairs are symmetric in color, and therefore also in
flavor.  Diquark condensates of this form break no
new symmetries, and therefore {\em must} arise
in the CFL phase~\cite{Alford:1998mk,Alford:1999pa}. 
However because the QCD interaction
is repulsive between quarks that are symmetric
in color these condensates are numerically
insignificant~\cite{Alford:1998mk,Alford:1999pa,Ruester:2004eg}.
To find which phases occur in realistic quark matter,
we must take into account the strange quark mass and
equilibration under the weak interaction, and impose
neutrality under the color and electromagnetic gauge symmetries.
The arguments that favor (\ref{condensate}) are unaffected
by these considerations, but there is no reason for
the gap parameters to be equal once $M_s\neq 0$.
Much previous 
work~\cite{Alford:1999pa,Schafer:1999pb,Buballa:2001gj,Alford:2002kj,Steiner:2002gx,Neumann:2002jm}
compared 
color-flavor-locked (CFL) phase (favored in the limit $M_s\to 0$
or $\mu\to\infty$), the two-flavor (2SC) phase 
(favored in the limit $M_s\to \infty$), and unpaired quark matter.
We gave a model-independent argument
in Ref.~\cite{Alford:2003fq}, however, that when the CFL phase is disrupted,
it cannot give way to either 2SC or unpaired quark matter. 
Above a critical $M_s^2/\mu$, we found that the CFL phase 
is replaced by a   new gapless CFL (gCFL) phase, not by 2SC quark
matter.  The defining (and eponymous) properties of the
gapless CFL phase arise in its dispersion relations,
not in its pattern of gap parameters.  However, it is
useful for orientation to list the patterns of 
gap parameters for all the phases we shall discuss:
\begin{eqnarray}
\De_3 \!\simeq\! \De_2 \!=\! \De_1 \!=\!\De_{CFL}
  &\ \ \ & \mbox{CFL} \label{CFL} \\
\De_3>0,\quad  \De_1 \!=\! \De_2 \!=\! 0
  &\ \ \ & \mbox{(gapless) 2SC} \label{2SC} \\
\De_2>0,\quad  \De_1 \!=\! \De_3 \!=\! 0
  &\ \ \ & \mbox{2SCus} \label{2SCus} \\
\De_3 > \De_2 >\De_1>0  &\ \ \ & \mbox{gapless CFL}\ . \label{gCFL}
\end{eqnarray}
The 2SCus phase, 
which was introduced in Ref.~\cite{Alford:2002kj},
must be analyzed for completeness because it and the 2SC phase
have the same free
energy when $M_s=0$, and to leading order in
$M_s$ if their respective nonzero
gap parameters have the same value~\cite{Alford:2002kj}.
However, we shall show in Section \ref{sec:g2SC}
that the 2SCus phase is never favored, and never
gapless.

In the remainder of this paper we construct the free energies
and solve the gap equations for the CFL, gapless CFL, 2SC, 
gapless 2SC~\cite{Shovkovy:2003uu,Huang:2003xd},
and 2SCus phases in an NJL model. We
show in detail how the CFL$\rightarrow$gCFL transition occurs
and detail the properties of the gCFL phase.  
The gCFL phase is a $\tilde Q$-conductor with a nonzero
electron density, and
these electrons and
the gapless quark quasiparticles make the low energy effective
theory of the gapless CFL phase and, consequently, its
astrophysical properties
qualitatively different from that of the CFL phase,
even though its $U(1)$ symmetries are the same.  
Both gapless
quasiparticles have quadratic dispersion relations at the
quantum critical point. For values of $M_s^2/\mu$ above the quantum critical
point, one branch has conventional linear dispersion relations
while the other branch remains quadratic, up to tiny corrections.
In order to evaluate the range of $M_s^2/\mu$ above the
critical point within which the gCFL phase remains favored,
we construct the 2SC and 2SCus phases and
reproduce the 2SC$\rightarrow$g2SC transition 
of Refs.\cite{Shovkovy:2003uu,Huang:2003xd}, 
here in neutral 3-flavor quark matter,
and show that in this context gCFL has a lower free energy
than (g)2SC(us).  We do not complete the study of mixed
phase alternatives, but we do eliminate all the most straightforward
possibilities everywhere in the gCFL regime in $M_s^2/\mu$ 
except very close to its upper end, where gCFL, g2SC and
unpaired quark matter have comparable free energies. 
At such large values of $M_s^2/\mu$, however, our pairing ansatz
is not sufficiently general to describe all the possibilities,
as we discuss in the concluding section of this paper.
Before turning to the model analysis, which we detail
in Section \ref{sec:model} and whose results 
we present in Section \ref{sec:results},
we conclude this introduction with a model-independent
discussion of color and electric neutrality in QCD
and with the model-independent argument of Ref.~\cite{Alford:2003fq}.

\subsection{Color and Electric Neutrality in QCD}

Stable bulk matter must be neutral under
all gauged charges, whether they are spontaneously broken or not.
Otherwise, the net charge density would create large electric fields,
making the energy non-extensive.
In the case of the electromagnetic gauge symmetry, this simply 
requires zero charge density.
In the case of the color gauge symmetry, the formal requirement is
that a chunk of quark matter should be a color singlet, i.e.,
its wavefunction should be invariant under a general color gauge
transformation. Color neutrality, meaning equality
in the numbers of red, green,
and blue quarks, 
is a less stringent constraint.
A color singlet state is also
color neutral, whereas the opposite is not necessarily true. However it has
been shown that 
the projection of a color neutral state
onto a color singlet state costs no extra free energy in the
thermodynamic limit~\cite{Amore:2001uf}.  Analyzing the consequences of the
requirement of color
neutrality therefore suffices for our purposes.

In nature, electric and
color neutrality are enforced by the dynamics of the electromagnetic
and QCD gauge fields, whose zeroth components serve as chemical potentials 
which take on values that enforce 
neutrality~\cite{Alford:2002kj,Gerhold:2003js}.  
Since we are
limiting ourselves to color neutrality and not color singletness we
have to consider only the $U(1)\times U(1)$ 
diagonal subgroup of the color
gauge group. This subgroup is generated by the 
diagonal generators 
$T_3=\diag(\half,-\half,0)$ and $T_8=\diag(\third,\third,-\twothirds)$
of the $SU(3)$ gauge group. Electromagnetism is generated by
$Q=\diag(\frac{2}{3},-\frac{1}{3},-\frac{1}{3})$ in flavor space (u, d, s). 
The zeroth components of the respective gauge fields serve as
chemical
potentials $\mu_3$ and $\mu_8$ coupled to $T_3$ and $T_8$ charges,
and as an electrostatic potential $\mu_e$ coupled to the
{\em negative} electric charge $Q$.  (We make this last choice
so that $\mu_e>0$ corresponds to a density of electrons, not
positrons.) The dynamics of the 
gauge potentials then require that the charge
densities, which are the derivatives of the free energy with respect to
the chemical potentials, must vanish:
\beq
\label{neutrality}
\ba{rcl}
Q = \phantom{-}\dsp\frac{\partial \Omega}{\partial\mu_e}&=& 0 \\[2ex]
T_3= -\dsp\frac{\partial \Omega}{\partial\mu_3}&=& 0 \\[2ex]
T_8= -\dsp\frac{\partial \Omega}{\partial\mu_8}&=& 0\ .
\ea
\eeq

A generic diquark condensate will be neither electrically nor
color neutral, so it will spontaneously break these gauge symmetries.
However it may be neutral under a linear combination
of electromagnetism and color. Indeed, any condensate of
the form (\ref{condensate}) is neutral with respect to
the ``rotated electromagnetism'' generated by 
$\tilde Q = Q -T_3 -\half T_8$, so $U(1)_{\tilde Q}$ is never broken.
This means that the corresponding gauge boson
(the ``$\Qtilde$-photon''), a mixture of the ordinary photon and
one of the gluons, remains massless. 
In both the CFL and gCFL phases,
the rest of the
$SU(3)_{\rm color} \times U(1)_Q$
gauge group is spontaneously broken, meaning
that the combination of the photon and gluons
orthogonal to the $\Qtilde$-photon, and all the
other gluons, become massive by the Higgs mechanism.

In an NJL
model with fermions but no gauge fields, as we shall
employ after pursuing model-independent arguments as far
as we can, one has to
introduce the chemical potentials $\mu_e$, $\mu_3$ and
$\mu_8$ ``by hand'' in order to enforce 
color and electric neutrality in the same way that gauge field
dynamics does
in QCD~\cite{Alford:2002kj}.

\subsection{Where does CFL pairing become unstable?}

We conclude this introduction with the
model-independent argument of Ref.~\cite{Alford:2003fq} that
determines the density at which the CFL phase becomes unstable.
The gap equations for the three $\De$'s will turn out to
be coupled, but we can, for example, 
analyze the effect of a specified $\De_1$ on the $gs$ and $bd$ quarks 
without reference to the other quarks. It turns out that
$gs$-$bd$ pairing is the first to break down, and this 
instability is what catalyzes the CFL$\rightarrow$gCFL transition.

The leading effect of $M_s$ is like a shift in the chemical potential
of the strange quarks, so the $bd$ and $gs$ quarks
feel ``effective chemical potentials'' 
$\mu_{bd}^{\rm eff} = \mu - \twothirds \mu_8$ and
$\mu_{gs}^{\rm eff} = \mu  + \third \mu_8 -\frac{M_s^2}{2\mu}$.
In the CFL phase, color neutrality
requires $\mu_8=-M_s^2/2\mu$, a result
that is model-independent to leading
order in $M_s^2/\mu^2$~\cite{Alford:2002kj,Steiner:2002gx}.
This result can be understood
as arising because CFL pairing itself 
enforces equality in the number of $rd$ and $gu$ quarks,
in the number of $bu$ and $rs$ quarks,
and in the number of $gs$ and $bd$ quarks~\cite{Rajagopal:2000ff},
but in order to achieve neutrality the number density
of $(rd,gu)$ quarks must be reduced relative to that 
of the $(bu,rs)$ and $(gs,bd)$ quarks,
and this requires a negative $\mu_8$.
Because of the negative $\mu_8$,
$\mu_{bd}^{\rm eff} - \mu_{gs}^{\rm eff} = M_s^2/\mu$ in the CFL
phase.
The CFL phase will be stable as long as the
pairing makes it energetically favorable to maintain equality of the
$bd$ and $gs$ Fermi momenta, despite their differing 
effective chemical
potentials \cite{Rajagopal:2000ff}. It becomes unstable when
the energy gained from turning a 
$gs$ quark near the common Fermi momentum into a $bd$ quark 
(namely $M_s^2/\mu$) exceeds the cost
in lost pairing energy $2\De_1$. 
Hence, the CFL phase is stable when~\cite{Alford:2003fq}
\beq
\frac{M_s^2}{\mu} < 2\De_{\rm CFL}\ .
\label{CFLstable}
\eeq
For lower density, i.e.~larger $M_s^2/\mu$, the CFL phase must be      
replaced by some new phase with unpaired $bd$ quarks.
One might naively expect this phase to be either neutral unpaired
quark matter
or neutral 2SC quark matter, but it is known that these have higher free
energy than CFL for 
$M_s^2/\mu < 4\De_{\rm CFL}$~\cite{Alford:2002kj,Steiner:2002gx}, 
so this new phase, which must have
the same free energy as CFL at the critical 
$M_s^2/\mu = 2\De_{\rm CFL}$, must be something else. In view of
its properties that are discussed in detail 
in Section \ref{sec:results}, we call it
gapless CFL (gCFL).

\section{Model and Approximations}
\label{sec:model}

We are interested in physics at non-asymptotic densities, and therefore
cannot use weak-coupling methods.  We are interested in physics at
zero temperature and high density, at which the fermion sign problem
is acute and the current methods of lattice QCD can therefore not be
employed.  For this reason, we need to introduce a model in which the
interaction between quarks is simplified, while still respecting the
symmetries of QCD, and in which the effects of $M_s$, $\mu_e$, $\mu_3$
and $\mu_8$ on CFL pairing can all be investigated.  The natural
choice is to model the interactions between quarks using a point-like
four-fermion interaction, which we shall take to have the quantum
numbers of single-gluon exchange.
We work in Euclidean space.
Our partition function
$\cal Z$ and free energy density $\Om$ are then defined by
\beq
\label{pathintegral}
\ba{rcl}
{\cal Z} &=&\dsp\e^{-\beta V \Omega}= {\cal N}\int {\cal D}\psibar{\cal D}\psi
  \exp\Bigl(\int {\cal L}(x)\,d^4x\Bigr) \\[4ex]
{\cal L}(x) &=&\dsp \psibar(i\feyn{\partial} + \feyn{\bm{\mu}} - \bm{M})\psi 
 -\frac{3}{8}G(\psibar\Gamma_\mu^A\psi)(\psibar\Gamma^\mu_A\psi)
\ea
\eeq
where the fields live in a box of volume $V$ and Euclidean time length
$\be=1/T$, and $\feyn{\bm{\mu}}=\bm{\mu}\gamma_4$. 
The interaction vertex has the 
color, flavor, and spin structure
of the QCD gluon-quark coupling, $\Gamma_\mu^A = \gamma_\mu T^A$. 
The mass matrix $\bm{M}=\diag(0,0,M_s)$ in flavor space.
The chemical potential $\bm{\mu}$ is a diagonal color-flavor 
matrix depending on
$\mu$, $\mu_e$, $\mu_3$ and $\mu_8$. The normalization of the
four-fermion coupling
$3G/8$ is as in the first paper in Ref.~\cite{reviews}.
In real QCD the ultraviolet modes decouple because of asymptotic freedom,
but in the NJL model we have to add this feature by hand, through
a UV momentum cutoff $\Lambda$ in the momentum integrals.
The model therefore has two parameters, the four-fermion coupling $G$
and the three-momentum cutoff $\Lambda$,  
but it is more useful to parameterize the interaction
in terms of a physical quantity, namely
the CFL gap parameter at $M_s=0$ at a reference chemical potential
that we shall take to be $500$ MeV. We shall call this
reference gap $\De_0$. 
We have checked that if we vary the cutoff $\Lambda$
by 20\% while simultaneously
varying the bare coupling $G$ so as to keep $\De_0$ fixed,
then our results change by a few percent at most.
All the results that we present are for $\Lambda=800$~MeV
and for a coupling strength chosen such that 
$\De_0=25$~MeV.

We now sketch the derivation of the free energy $\Omega$ obtained
from the Lagrangian (\ref{pathintegral}) upon making
the ansatz (\ref{condensate}) for the diquark
condensate and working in the mean field approximation.
More sophisticated derivations exist in the literature~\cite{reviews},
but since we are 
assuming that the only condensate is of the form 
(\ref{condensate}) we simply
Fierz transform the interaction to yield products of
terms that appear in (\ref{condensate}), and discard
all the other terms that arise in the Fierz transformed
interaction which would anyway vanish after making the
mean field approximation.
This yields
\beq
{\cal L}_{\rm int} = \frac{G}{4} \sum_\eta (\bar\psi P_\eta \bar\psi^T)
(\psi^T \bar P_\eta \psi)\ ,
\label{fierzed}
\eeq
where 
\beq
\left(P_\eta\right)^{\al\be}_{ij} = C\ga_5 \eps^{\al\be \eta} \eps_{ij\eta}
\quad\mbox{(no sum over $\eta$)} 
\label{Pdefinition}
\eeq
and $\bar P_\eta = \ga_4 P^\dag_\eta \ga_4$. 
The index $\eta$ labels the pairing channel:
$\eta=1,2,$ and $3$ correspond to $d$-$s$ pairing,
$u$-$s$ pairing, and  $u$-$d$ pairing.
The overall coefficient in (\ref{fierzed}) is the product
of the $3G/8$ in (\ref{pathintegral}) and factors
of $-1$, $4/3$, and $-1/2$ from Fierz transformations in
Dirac, color and flavor space, respectively.

Next, for each channel we introduce a complex
scalar field $\phi_\eta$ whose expectation value will be
$\De_\eta$, the strength of the pairing in the $\eta$ channel,
 and bosonize the four-fermion interaction
via a Hubbard-Stratonovich transformation. The interaction
Lagrangian then becomes
\beq
{\cal L}_{\rm int} =\half (\bar\psi P_\eta \bar\psi^T)\phi_\eta
+ \half \phi_\eta^* (\psi^T \bar P_\eta \psi) 
- \frac{\phi_\eta^*\phi_\eta}{G}\ ,
\eeq
where here and henceforth repeated $\eta$'s are summed and
where it is understood that 
we are now integrating over the $\phi_\eta$ as well as $\psi$ and $\bar \psi$
in the functional integral \eqn{pathintegral}.
The functional integral is now quadratic in the quark fields, so
the fermionic function integral can be performed.
Since there are terms in the action that
can violate quark number, we must use Nambu-Gorkov spinors
\beq
\Psi = \left(\ba{c}\psi(p)\\ \bar\psi^T(-p)\ea\right),\quad
\bar\Psi = \Bigl( \bar\psi(p)\ \psi^T(-p) \Bigr)
\eeq
and the full Lagrange density becomes
\beq
{\cal L} = \frac{1}{2}\bar\Psi \frac{S^{-1}}{T}\Psi 
- \frac{\phi_\eta^*\phi_\eta}{G}
\eeq
where the inverse full propagator is
\beq
\label{invprop}
S^{-1}(p) = \left( \ba{cc}
\feyn{p} + \feyn{\mu}-\bm{M} & P_\eta\phi_\eta \\
\bar P_\eta\phi^*_\eta &  (\feyn{p} - \feyn{\mu}+\bm{M})^T \ea \right)\ .
\eeq
We now integrate over the fermionic fields to obtain the effective
potential for the scalar fields. We also make the mean-field approximation,
neglecting fluctuations
in the scalar fields and setting $\phi_\eta$ to its 
expectation value $\De_\eta$. 
The result is
\beq
{\cal Z}= \left[\Det\frac{S^{-1}(i\omega_n,p)}{T}\right]^{1/2}
\exp\left(-\frac{V}{T}\frac{\Delta_\eta\Delta_\eta}{G}\right)
\eeq
and hence
\beq
\Omega
=
-T\sum_{n}\int\!\frac{d^3p}{(2\pi)^3}\frac{1}{2}
 {\rm Tr}\log\Bigl(\frac{1}{T}S^{-1}(i\omega_n,p)\Bigr)
+\frac{\Delta_\eta\Delta_\eta}{G}\ ,
\eeq 
where $\omega_n=(2n-1)\pi T$ are the 
Matsubara frequencies. We do the Matsubara summation using
the identity
\beq
T \sum_{n}\ln\Bigl(\frac{\omega_n^2+\varepsilon^2}{T^2}\Bigr)
 =|\varepsilon| + 2T\ln(1+e^{-|\varepsilon|/T})\ .
\label{matsubara}
\eeq
In the limit of zero temperature only the first term from the 
right hand side survives, leading to the result
\beq
\label{omega}
\ba{rcl}
\Omega &=& \dsp - \frac{1}{4\pi^2}
  \int_0^\Lambda p^2 \sum_j|\varepsilon_j(p)|\, dp \\[2ex]
  &&\dsp + \frac{1}{G}(\De_1^2 + \De_2^2 + \De_3^2)
  - \frac{\mu_e^4}{12\pi^2}\ ,
\ea
\eeq
where the electron contribution is included, and
$\varepsilon_j(p)$ are the dispersion relations of the
quasiquarks, i.e.~the values of the energy at which
the propagator diverges:
\beq
\label{disprel}
\det S^{-1}(i\varepsilon_j(p),p)=0 \ .
\eeq
$S^{-1}$ is a $72\times 72$ matrix, but because what
occurs in the identity (\ref{matsubara}) is
the combination $\omega^2+\varepsilon^2$, the sum in 
(\ref{omega}) is understood to run over 36 roots. 
(This can be seen as removing the doubling of degrees of
freedom introduced by using the Nambu-Gorkov formalism.)
In the specific cases where our general ansatz becomes 2SC or CFL
pairing, our expression (\ref{omega}) for the free energy,
and in particular the coefficient of the $\De^2$ term,
agrees with
the expressions obtained by other methods~\cite{reviews} that do not
involve Fierz transformations.

In our numerical evaluation, we omit the antiparticle modes: exciting
them costs of order $2\mu$ and they therefore do not play an important
role in the physics. This is discussed in more detail below.
Neglecting the antiparticles leaves
us with only 18 roots of (\ref{disprel}) to sum over in (\ref{omega}).
These correspond to 9 different dispersion
relations describing the quasiparticles of
differing color and flavor, each doubly
degenerate due to spin.

A stable, neutral phase must 
minimize the free energy (\ref{omega}) 
with respect to variation of the three gap parameters 
$\Delta_1$, $\Delta_2$, $\Delta_3$, meaning it must satisfy
\beq
\label{gapeqns}
\dsp\frac{\partial\Omega}{\partial\Delta_1} = 0, \qquad 
\dsp\frac{\partial\Omega}{\partial\Delta_2} = 0, \qquad
\dsp\frac{\partial\Omega}{\partial\Delta_3} = 0\ ,
\eeq
and it must satisfy the three neutrality conditions
\eqn{neutrality}.
The gap equations (\ref{gapeqns}) 
and neutrality equations \eqn{neutrality} 
form a system of six coupled integral equations 
with unknowns the three gap parameters and $\mu_3, \mu_8$ and $\mu_e$.

We must now find the dispersion relations  $\varepsilon_j(p)$,
determined by the
zeroes of $\det S^{-1}$ which is specified by (\ref{disprel}), 
(\ref{invprop}) with $\phi_\eta$ replaced by $\Delta_\eta$, 
and (\ref{Pdefinition}), then evaluate the free energy $\Omega$
using (\ref{omega}), and then solve the 
six simultaneous equations \eqn{neutrality} and
\eqn{gapeqns}. Before carrying this calculation
through, however, 
we first make a number of simplifying approximations
within the expression for $\det S^{-1}$.
\begin{enumerate}
\setlength{\itemsep}{-\parsep}\setlength{\topsep}{0ex}
\item We neglect
contributions to the condensate that are symmetric in
color and flavor: these are known to be present and 
small \cite{Alford:1998mk,Alford:1999pa,Ruester:2004eg}.
\item We treat the up and down quarks as massless, which is a legitimate 
approximation in the high density regime, and we treat the constituent 
strange quark mass $M_s$ as a parameter, rather than solving
for an $\langle \bar s s \rangle$ condensate.  The latter approximation
should be improved upon, along the lines of Ref.~\cite{Steiner:2002gx}.
\item We incorporate
$M_s$ only via its leading effect, namely
as a shift $-M_s^2/2\mu$ in the chemical potential
for the strange quarks. 
This 
approximation neglects the difference between the strange
and light quark Fermi velocities, whose effects are known
in other contexts to be small~\cite{Kundu:2001tt}.
The approximation
is controlled by the smallness of $M_s^2/\mu^2$. 
For this reason, in all the results
that we plot we shall work at $\mu=500$~MeV and 
choose a coupling such that the CFL gap
at $M_s=0$ is
$\Delta_0=25$~MeV.  We expect the CFL pairing to break down
near $M_s^2\approx 2\mu\De_0$, and choosing $\De_0=\mu/20$ 
ensures that this occurs where $M_s^2/\mu^2\sim 1/10$, 
meaning that we can trust our results well into the 
gapless CFL phase~\cite{kenji:private}.  If, instead,
we choose a larger $\De_0$, as in Ref.~\cite{Ruester:2004eg},
we find that our results become markedly more $\Lambda$-dependent,
which is a good diagnostic for model-dependence.  
\item We work to leading
nontrivial order in $\De_1$, $\De_2$, $\De_3$, $\mu_e$, $\mu_3$ 
and $\mu_8$. This should be a good approximation, as all these
quantities are small compared to $\mu$. 
\item We neglect the anti-particles. This simplifies the numerics
by discarding physically unimportant degrees of freedom,
but one must be cautious with this truncation.
It introduces cutoff-dependent terms in our free
energy, including some that depend on the chemical potential and
therefore introduce cutoff-dependence in the corresponding charges.
For our purposes this is not important, firstly because we always
present free energy differences relative to neutral unpaired quark
matter, and secondly because we only care about electric and color
charges that have zero trace over all fermion species, and for these
the cutoff dependence cancels out. However, a non-traceless charge
like baryon number would have an incorrect cutoff-dependent value when
calculated in this approximation.
\item We ignore meson condensation in both the CFL and gCFL phases.
\end{enumerate}
We expect that these approximations
have quantitative effects, but none preclude a qualitative
understanding of the new phase we shall describe.

We now give the explicit expression for $\det S^{-1}$, after having
implemented the approximations above.
As described in section \ref{sec:ansatz} we use a color-flavor basis
in which the gap matrix (\ref{blocks}) is conveniently block-diagonal.
Since the
chemical potential and mass are diagonal in color and flavor,
the full inverse propagator \eqn{invprop}
is then also block-diagonal in color-flavor space.
This means we can break the 
determinant in equation \eqn{disprel}
into four more manageable pieces:
\beq
\det S^{-1}(p_0,p) = \left(\Drugdbs \Drdgu \Drsbu \Dgsbd\right)^2\ .
\label{determinantinpieces}
\eeq
We find
that the $2\times 2$ determinants are
\beq
\ba{rcl}
\Drdgu &=& 16\mu^4((\mu_{rd}-p-ip_0)(\mu_{gu}-p+ip_0)+\Delta_3^2)\\[-1ex]
       &&  ((\mu_{rd}-p+ip_0)(\mu_{gu}-p-ip_0)+\Delta_3^2), \\[2ex]
\Drsbu &=& 16\mu^4((\mu_{rs}-p-ip_0)(\mu_{bu}-p+ip_0)+\Delta_2^2)\\[-1ex]
       && ((\mu_{rs}-p+ip_0)(\mu_{bu}-p-ip_0)+\Delta_2^2), \\[2ex]
\Dgsbd &=& 16\mu^4((\mu_{bd}-p-ip_0)(\mu_{gs}-p+ip_0)+\Delta_1^2)\\[-1ex]
       && ((\mu_{bd}-p+ip_0)(\mu_{gs}-p-ip_0)+\Delta_1^2),
\ea
\eeq
and the $3\times 3$ determinant is
\beq
\ba{lll}
\Drugdbs =& a(2\mu)^6\Bigl( 
  & b(de-\De_1^2)(cf-\De_1^2)-cdf\De_2^2 \\[-3ex]
   && +d\De_1^2\De_2^2-def\De_3^2+f\De_1^2\De_3^2\,\Bigr) \\[1ex]
+(2\mu)^6\Bigl( & \multicolumn{2}{l}{ 
      \De_3^2(de\De_2^2-4\De_1^2\De_2^2+ef\De_3^2)}\\
  & \multicolumn{2}{l}{ +c(d\De_2^4+f\De_2^2\De_3^2)} \\
 &  \multicolumn{2}{l}{ +b(e\De_1^2\De_3^2
     +c(\De_1^2\De_2^2-de\De_2^2-ef\De_3^2))\,\Bigr)}
\ea
\eeq
where for compactness we assign
\beq
\ba{rcl}
 a &=& \mu_{ru}-p+ip_0 \\[1ex]
 b &=&  -\mu_{ru}+p+ip_0 \\[1ex]
 c &=&  \mu_{gd}-p+ip_0 \\[1ex]
 d &=&  -\mu_{gd}+p+ip_0 \\[1ex]
 e &=&  \mu_{bs}-p+ip_0 \\[1ex]
 f &=&  -\mu_{bs}+p+ip_0
\ea
\eeq
and where we have dropped the superscript on the 
``effective quark chemical potentials'', given by
\beq
\ba{rcl}
 \mu^\mathrm{eff}_{ru}&=& \mu-\twothirds\mu_e+\half\mu_3+\third\mu_8,\\[1ex]
 \mu^\mathrm{eff}_{gd}&=& \mu+\third\mu_e-\half\mu_3+\third\mu_8,\\[1ex]
 \mu^\mathrm{eff}_{bs}&=& 
\mu+\third\mu_e-\twothirds\mu_8-M_s^2/(2\mu),\\[3ex]

 \mu^\mathrm{eff}_{rd}&=& \mu+\third\mu_e+\half\mu_3+\third\mu_8,\\[1ex]
 \mu^\mathrm{eff}_{gu}&=& \mu-\twothirds\mu_e-\half\mu_3+\third\mu_8,\\[3ex]

 \mu^\mathrm{eff}_{rs}&=& 
\mu+\third\mu_e+\half\mu_3+\third\mu_8-M_s^2/(2\mu),\\[1ex]
 \mu^\mathrm{eff}_{bu}&=& \mu-\twothirds\mu_e-\twothirds\mu_8,\\[3ex]

 \mu^\mathrm{eff}_{gs}&=& 
\mu+\third\mu_e-\half\mu_3+\third\mu_8-M_s^2/(2\mu),\\[1ex]
 \mu^\mathrm{eff}_{bd}&=& \mu+\third\mu_e-\twothirds\mu_8\ . \\[1ex]

\ea
\eeq
These expressions explicitly show how we treat
the strange quark mass as a shift in the chemical 
potential of the strange quarks.
In evaluating these determinants, we have 
extensively used the identity 
\begin{displaymath}
\det\left( \begin{array}{cc}
A & B \\
C & D 
\end{array}\right)=\det(A)\det(D-CA^{-1}B)
\end{displaymath}
for the determinant of a block matrix.

The numerical task is now explicit.  We find the 
quasiparticle dispersion relations $\varepsilon(p)$ by finding
the zeros of (\ref{determinantinpieces}), viewed
as a polynomial in $p_0$. 
We then perform the integral
in (\ref{omega}) numerically, and obtain $\Omega$.
We evaluate the partial derivatives of the free energy
required in the neutrality
conditions (\ref{neutrality}) and 
the gap equations (\ref{gapeqns}) numerically as finite
differences, with differences $0.1$~MeV in the relevant 
chemical potential or gap parameter.

As a check, we have also done the calculation of
$\Omega$ and its partial derivatives by 
evaluating both the $p$ and $p_0$ integrals 
numerically, never writing the latter as a Matsubara sum.
In this alternative calculation, we were able to
evaluate the partial derivatives in (\ref{neutrality})
and (\ref{gapeqns}) analytically.

The $\Omega$ that we obtain is cutoff dependent, but its 
partial derivatives (\ref{neutrality})
and (\ref{gapeqns}) are insensitive to variations
in $\Lambda$ in the sense described above, namely
as long as the coupling is changed to keep $\De_0$ fixed
upon variation in $\Lambda$, and as long as $\Lambda$ is kept
well above $\mu$.  Furthermore, we 
are only ever interested in free energy differences between
phases.
When we evaluate the differences
between the $\Omega$ for unpaired, (g)CFL, and (g)2SC quark
matter, we find that all such free energy differences are
insensitive to the cutoff, as they should be since these differences
all reflect physics near the Fermi surfaces.
Because we are only interested in free energy differences,
in evaluating $\Omega$ we make the numerical integral 
better behaved by subtracting
the appropriate expression for
neutral unpaired quark matter
within the integrand.

The solutions of the system of gap and neutrality equations
depend on three parameters: $\mu$, $M_s$ and
$\De_0$. Our purpose is to understand the effect of $M_s$
on CFL pairing, and these effects are controlled by
the relative size of $M_s^2/\mu$ and the gap parameters $\De_i$,
whose overall magnitude is set by $\De_0$.
It is therefore better to think of the
three parameters in the problem
as  $\mu$, $M_s^2/\mu$ and $\De_0$.
In compact stars,
$\mu$ increases and $M_s$ presumably decreases, 
meaning that $M_s^2/\mu$ decreases as one approaches
the center of the star. For simplicity, we set the
overall energy scale in our calculation by fixing
$\mu=500$~MeV, which is reasonable for the center of a neutron star,
and vary $M_s$ in order to vary $M_s^2/\mu$.
We have confirmed that as long as we choose a $\De_0$ that
is small enough that the transition (\ref{CFLstable}) occurs where
$M_s^2/\mu^2$ corrections are under 
control, this transition occurs very close to
$M_s^2/\mu\approx 2 \De$, 
where $\De$ is
gap parameter on the CFL side of the transition.  
The authors of Ref.~\cite{Ruester:2004eg} have confirmed 
that this result continues to be valid
even for $\De_0$ as large as $100$~MeV, where the approximations
are not as well controlled.
We quote results only for $\De_0=25~\MeV$, 
which is within the plausible range of values that 
$\De_0$ may take in nature~\cite{reviews}
and for which our calculation is clearly under control.
Although we have obtained our results by varying $M_s$ at
fixed $\mu$,
we typically quote results in terms of the important combination 
$M_s^2/\mu$.

\section{Results}
\label{sec:results}

\subsection{Domain where gCFL is favored}

\begin{figure}[t]
\begin{center}
\includegraphics[width=0.48\textwidth]{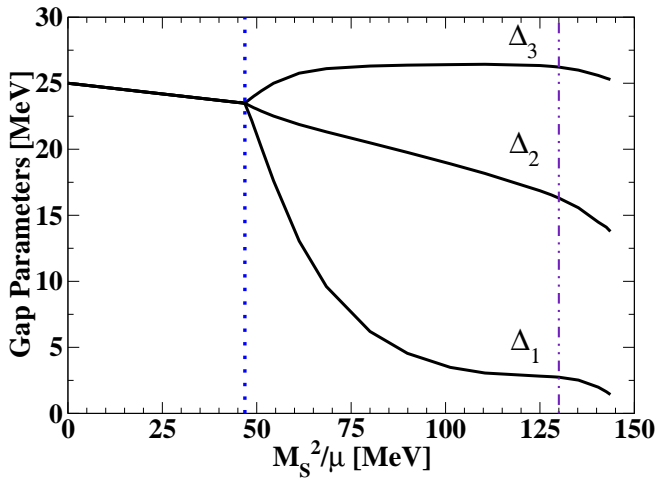}
\end{center}
\vspace{-0.25in}
\caption{
Gap parameters $\De_3$, $\De_2$, and $\De_1$
as a function of $M_s^2/\mu$ for $\mu=500$~MeV, in a model
where $\De_0=25$~MeV (see text). 
At $M_s^2/\mu\approx 47.1~\MeV$ (vertical dotted line)
there is a continuous phase
transition between the CFL phase 
and a phase that we shall identify
below as 
the gapless CFL phase. 
We find gapless CFL phase solutions
up
to $M_s^2/\mu\approx 144$~MeV.
But, we 
shall see in Fig.~\ref{fig:energy} that
above $M_s^2/\mu\approx 130~\MeV$ 
(which we denote here with
a vertical dash-dotted line)
unpaired quark matter has 
a lower free energy than the gapless CFL phase.
}
\label{fig:gaps}
\end{figure}

\begin{figure}[t]
\begin{center}
\includegraphics[width=0.48\textwidth]{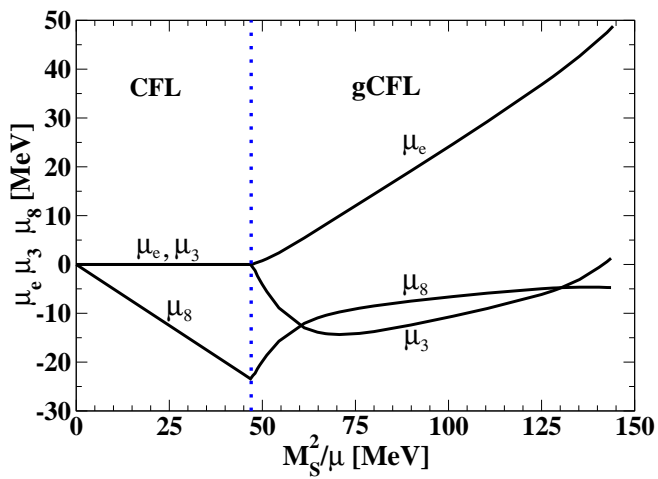}
\end{center}
\vspace{-0.25in}
\caption{
Chemical potentials $\mu_e$, $\mu_3$ and $\mu_8$ 
as a function of $M_s^2/\mu$ in
the CFL/gCFL phase for the same parameters
as in Fig.~\ref{fig:gaps}.  The effects of electrons on the
free energy have been included in the calculation, as
will be discussed in more detail below.  We see that
the gapless CFL phase has $\mu_e>0$, meaning that it
has a nonzero density of electrons.  Perhaps the most
physically relevant order parameter for the CFL/gCFL
phase transition is the electron number density
$n_e\sim \mu_e^3$.
}
\label{fig:mue_mu3_mu8}
\end{figure}

\begin{figure}[t]
\begin{center}
\includegraphics[width=0.49\textwidth]{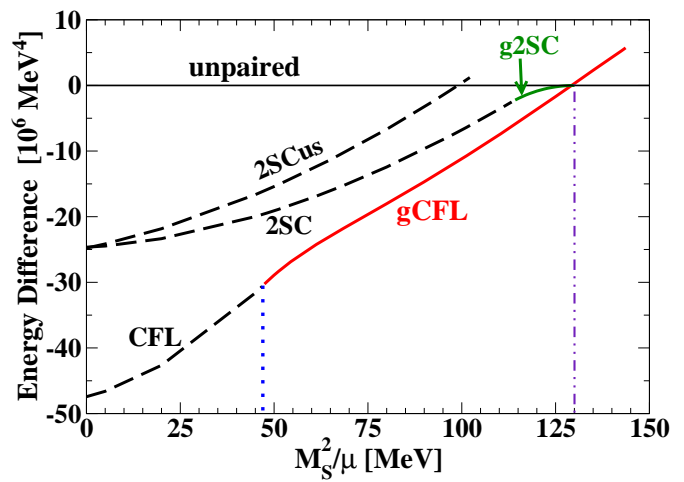}
\end{center}
\vspace{-0.25in}
\caption{
Free energy of the CFL/gCFL phase, relative to that
of neutral noninteracting quark matter and that
of the 2SC/g2SC and 2SCus phases, discussed in
Section \ref{sec:g2SC}.
There is a CFL$\to$gCFL transition at $(M_s^2/\mu)_c\approx 47.1~\MeV$,
(vertical dotted line), at which the free energy and its slope are
continuous, indicating that the transition is not first order.
If we neglect the possibility of other phases (for example
g2SC) we would conclude from this figure that
there is a first order transition gCFL$\to$unpaired
at $M_s^2/\mu\approx 130~\MeV$ (vertical dash-dotted line).
}
\label{fig:energy}
\end{figure}

In Figs.~\ref{fig:gaps} and \ref{fig:mue_mu3_mu8}, 
we show the gap parameters and chemical potentials as a function
of $M_s^2/\mu$, for 
$\De_0=25~\MeV$. Fig.~\ref{fig:energy} shows the free energy.
We see
a continuous phase transition occurring at a critical
$M_s^c$ that, in our model
calculation with $\mu=500$~MeV, lies between $M_s=153~\MeV$ and 
$M_s=154~\MeV$, i.e.~at $(M_s^2/\mu)_c \approx 47.1~\MeV$.
This agrees exceedingly well with the expected value $2\Delta$
from \Eqn{CFLstable}, since on the CFL side
of the transition $\De_1=\De_2=\De_3=23.5$ MeV.
For $M_s^2/\mu < (M_s^2/\mu)_c$, the CFL phase is favored,
with all three gaps equal to each other within our approximations.
If we improve upon our 
approximate treatment of $M_s$, we expect $\De_1=\De_2$ with
these gap parameters slightly smaller than $\De_3$, because
$\De_1$ and $\De_2$ describe pairing between quarks with
differing Fermi velocities, an effect of $M_s$ that we 
are neglecting because it is known to be small in other
contexts~\cite{Kundu:2001tt}. (Indeed, it proves
to be a few percent effect also in the 
present context~\cite{kenji:private}.)

For $M_s^2/\mu < (M_s^2/\mu)_c$ our results agree with the small-$M_s$
expansion of Ref.~\cite{Alford:2002kj}, where
$\mu_3=\mu_e=0$ and
$\mu_8=-M_s^2/(2\mu)$ and the free energy is
\begin{eqnarray}
\Omega_{\mbox{\scriptsize\begin{tabular}{l}neutral\\ CFL\end{tabular}}}
  &=& - \frac{3 \mu^4}{4\pi^2} 
+ \frac{3 M_s^2 \mu^2}{4\pi^2} 
- \frac{1- 12\log(M_s/2\mu)}{32\pi^2}M_s^4\nonumber\\
& &\qquad - \frac{3\Delta^2\mu^2}{\pi^2}\nonumber\\
&=& \Omega_{\mbox{\scriptsize\begin{tabular}{l}neutral\\ unpaired\end{tabular}}}+ \frac{ 3 M_s^4 - 48\Delta^2\mu^2}{16\pi^2}\ .
\label{naiveCFL}
\end{eqnarray}

As the density decreases (i.e.~as $M_s^2/\mu$ increases) through
the CFL$\to$gCFL transition, the gap parameters split apart,
with $\Delta_3$  increasing slightly 
and $\Delta_2$ and $\Delta_1$ dropping significantly, with
$\Delta_1$ dropping faster than $\Delta_2$.

We have verified that $M_s^2/\mu\De$ is the relevant dimensionless quantity
by changing the coupling strength, i.e.~picking a
different $\Delta_0$ (gap at $M_s=0$). The critical 
point $(M_s^2/\mu)_c$ changes as
predicted by \eqn{CFLstable}. Furthermore we checked 
the robustness of our results upon
variation of the cutoff $\Lambda$, observing changes of only a few
percent in the value of $M_s^2/\mu\De$ at the transition
upon changing $\Lambda$ by up to 20\% while keeping
$\De_0$ fixed.

Fig.~\ref{fig:energy} confirms that the slope of the free energy is continuous
at the CFL/gCFL transition,
indicating that it is not first order.  We have not determined
the order of the transition, because evaluating higher derivatives
of the free energy with respect to $M_s^2/\mu$ is not numerically
feasible.  The most physically relevant order parameter
is the electron density $n_e\sim \mu_e^3$, which is of course
equal in magnitude to the electric charge density 
of the quarks.
This increases above the transition like
$n_e\sim \left[(M_s^2/\mu)-(M_s^2/\mu)_c \right]^3$, suggesting
a fourth order phase transition! This argument neglects the
small electron mass and, furthermore, it neglects the fact that,
as we shall see in \Eqn{breach}, there is also a nonzero number density 
of neutral unpaired quark quasiparticles that grows like
$\left[(M_s^2/\mu)-(M_s^2/\mu)_c \right]^{1/2}$. Although
because these unpaired quasiparticles are neutral
they are less important phenomenologically, this does suggest that the
transition is second order, as in the analysis of Ref.~\cite{Liu:2004mh}.

If we had used a simpler ansatz in which the gap parameters were
constrained to one value $\De\equiv\De_1=\De_2=\De_3$, then 
the CFL phase would have remained artificially stable above the 
critical value of $M_s^2/\mu$. 
From Eq.~(\ref{naiveCFL}),  its free energy would be higher
than that of gCFL, rising to equality with that 
of unpaired quark matter at a value of $M_s^2/\mu$ around $90$~MeV.
(The precise value depends on how the common $\De$ changes with $M_s$.)

Of course, we actually used the more general
ansatz \eqn{condensate} that allows the $\De$'s to differ.
We found that the CFL phase becomes unstable and is
replaced by the gCFL phase, in which the gaps have very different values,
so the simplified analysis of Ref.~\cite{Alford:2002kj} does not apply.
The free energy of the gCFL phase crosses that of unpaired quark matter 
at $M_s^2/\mu\approx 130~\MeV$.
This phase transition is first order,
and we are able to follow the metastable gCFL phase up to
$M_s^2/\mu=144~\MeV$ where, as we shall explain below, it ceases
to be a solution.

\begin{figure}[t]
\begin{center}
\includegraphics[width=0.48\textwidth]{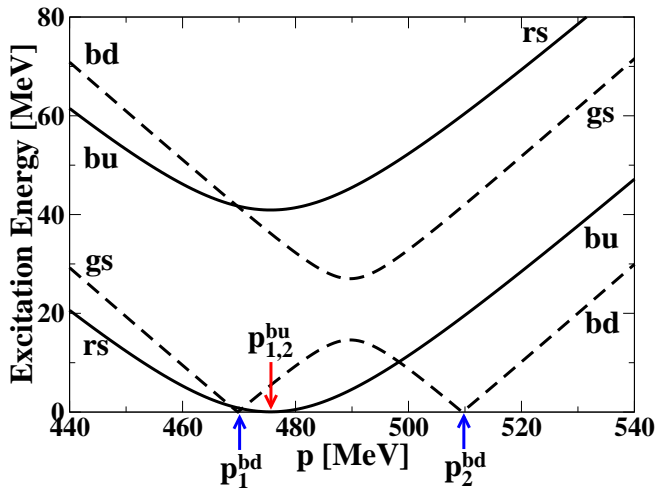}
\end{center}
\vspace{-0.2in}
\caption{Dispersion relations at $M_s^2/\mu=80~\MeV$
(with $\mu$ and $\De_0$ as in previous figures)
for the quasiparticles that are linear combinations of
$gs$ and $bd$ quarks (dashed lines)
and for the quasiparticles
that are linear combinations of $bu$ and $rs$ quarks (solid lines).
There are gapless $gs$-$bd$ modes at
$p_1^{bd}=469.8~\MeV$ and $p_2^{bd}=509.5~\MeV$, which are
the boundaries of the 
``blocking''~\cite{Alford:2000ze,Bowers:2001ip} 
or 
``breached pairing''~\cite{Gubankova:2003uj} region
wherein there are unpaired $bd$ quarks and no $gs$ quarks.
One $bu$-$rs$ mode is gapless at $p=475.6~\MeV$
with an almost exactly quadratic dispersion relation
that we shall discuss below. 
}
\label{fig:disprel}
\end{figure}

\subsection{The nature of the gCFL phase}

\begin{figure}[t]
\begin{center}
\includegraphics[width=0.48\textwidth]{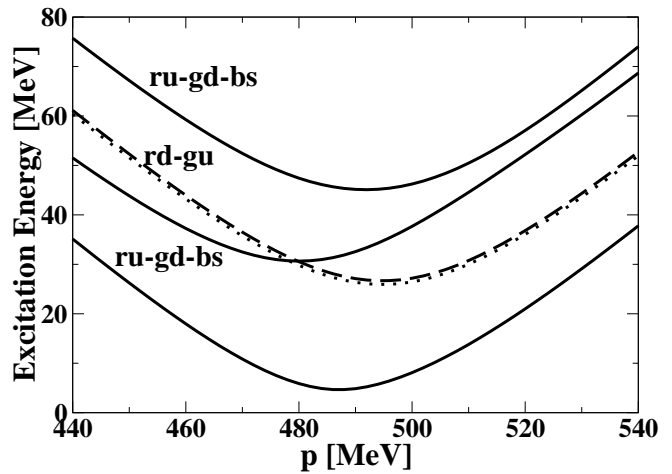}
\end{center}
\vspace{-0.25in}
\caption{Dispersion relations at $M_s^2/\mu=80~\MeV$
for the two quasiparticles that are linear combinations
of $rd$ and $gu$ quarks (dashed and dotted), and for 
the three quasiparticles that are linear
combinations of $ru$, $gd$ and $bs$ quarks (solid).
These five quark 
quasiparticles all have gaps throughout
the CFL and gCFL phases.
}
\label{fig:dispersion_5modes}
\end{figure}

Up to this point we have not justified
our use of the name ``gapless CFL'' for
the new phase that replaces the CFL phase at $M_s^2/\mu\gtrsim 2\De$.
We have given model-independent arguments to expect that it will
contain unpaired  $bd$ quarks, but
now we describe its properties in more detail.
In calculating the free energy \eqn{omega} of the Cooper-paired
quark matter we automatically obtain the quasiquark
dispersion relations \eqn{disprel}, so we can 
see what gapless modes exist. These modes are important because,
at the temperatures $T\lesssim \keV$ characteristic of neutron
stars, only the lightest modes will contribute to transport
properties.

In Fig.~\ref{fig:disprel} we show the dispersion relations for
the $rs$-$bu$ and $gs$-$bd$ $2\times 2$ blocks in the quasiquark
propagator, at $M_s^2/\mu=80~\MeV$. We see immediately that there
are gapless modes in both blocks, justifying our name for this phase.
Before moving on to a detailed discussion of the physical properties
of the gCFL phase, we should note that the phenomenon of
gapless superconductivity is well known, at least theoretically.
It was first suggested by Sarma \cite{Sarma} who worked in a context much like
our $gs$-$bd$ block in isolation, and found that the gapless
superconducting phase is never stable.   Alford, Berges
and Rajagopal found a metastable
gapless color superconducting phase
in Ref. \cite{Alford:1999xc}, but this phase was neither electrically
nor color neutral.   
The key observation was made by
Shovkovy and Huang \cite{Shovkovy:2003uu}, who discovered that
when the constraints of electric and color neutrality
are imposed on the 2SC phase in two-flavor QCD, there
are regions of parameter space where a gapless color
superconducting phase is stable.  
Following their
nomenclature (they described a ``gapless 2SC phase'')
we refer to the phase that we find
above $M_s^c$ as the
``gapless color-flavor locked phase''~\cite{Alford:2003fq}.
     
Gapless two-flavor color
superconductivity was also studied in Ref.~\cite{Gubankova:2003uj},
building upon prior work done in a cold atom context~\cite{Liu:2002gi}.
These authors analyzed pairing between a heavy and a light quark,
akin to $gs$ and $bd$, in the case in which
the $gs$ quarks are nonrelativistic.   They find that a
gapless phase (they describe the blocking region as
a region in which pairing is ``breached'') is stable
if the relative density of the two species is held fixed.

Note that in our three-flavor calculation, both the gap equations
(\ref{gapeqns}) and the neutrality conditions (\ref{neutrality})
couple all nine quarks.  Although the single particle dispersion
relations can be analyzed for the $gs$ and $bd$ quarks in isolation,
and are qualitatively similar to those obtained in
Refs.~\cite{Shovkovy:2003uu,Gubankova:2003uj} in a two-flavor setting,
the implications of neutrality are more subtle in our three-flavor
context as we shall explain below.

Each of the dispersion relations in Figs.~\ref{fig:disprel}
and \ref{fig:dispersion_5modes}  describes
an excitation with well-defined $\tilde Q$, although
the sign of $\tilde Q$ changes at momenta where
the dispersion relation is gapless.
Beginning
with an example with no gap, 
the upper solid curve in Fig.~\ref{fig:disprel} describes excitations that
are linear combinations of $rs$ particles and $bu$ holes,
both with $\tilde Q=-1$.  The lower dashed curve in Fig.~\ref{fig:disprel} has
clearly visible momenta $p_1^{bd}$ and $p_2^{bd}$
where it is gapless, so we use this as an example of
``sign change'' even though it describes 
$\tilde Q=0$ quasiparticles: 
to the left of $p_1^{bd}$, it describes $gs$-holes
with a very small admixture of $bd$ particles; 
to the right of $p_2^{bd}$, it describes $bd$ particles
with a very small admixture of $gs$ holes; but, between
$p_1^{bd}$ and $p_2^{bd}$ it describes excitations that
are superpositions of $bd$ holes and $gs$ particles.

In the CFL phase, once we take into account
the explicit symmetry breaking 
introduced by the 
strange quark mass and electromagnetism,
the unbroken symmetry is
reduced from the diagonal 
$SU(3)_{L+R+c}$ to
$U(1)_{\tilde{Q}}\times U(1)$~\cite{Son:1999cm}. 
The last $U(1)$ corresponds to 
``color + flavor hypercharge'' and may be
spontaneously broken by meson condensation~\cite{Bedaque:2001je}.
The gapless CFL phase has the same symmetry as the CFL phase,
and it will therefore be interesting to investigate the possibility of
meson condensation in the gCFL phase. The effective theory
for the Goldstone bosons alone will have the same form as in
the CFL phase, albeit with new contributions to their masses
coming from the differences between the values of the
three $\De_i^2$.  And, furthermore, the gapless quasiparticles
must be included in the low energy effective theory.

\subsubsection{Dispersion relations, gapless modes, and neutrality}

As will soon become clear, the $3\times 3$ block in the pairing pattern
\eqn{blocks} plays a minor role: its quasiparticles are always
gapped, so we mainly discuss the three $2\times 2$ blocks.
In general, when two species of massless quarks 
undergo $s$-wave pairing with gap parameter $\De$,
the dispersion relations of the two resulting quasiparticles are
\begin{equation}
E(p) = \Bigl| \de\mu \pm \sqrt{ (p-\bar \mu)^2 + \De^2} \Bigr|
\label{disprel2}
\end{equation}
where the individual chemical potentials of the quarks
are $\bar\mu \pm \de \mu$. As long as the chemical potentials
pulling the two species apart are not too strong,
Cooper pairing occurs at all momenta:
\begin{equation}
\mbox{pairing criterion:}\quad |\de\mu|<\De\ .
\label{pairing}
\end{equation}
However when this condition is violated
there are gapless ($E=0$) modes at momenta
\begin{equation}
p_{\rm gapless} = \bar\mu \pm \sqrt{\de\mu^2-\De^2}
\label{blocking}
\end{equation}
and there is no pairing in the ``blocking'' or ``breached pairing''
region between these momenta
\cite{Alford:2000ze,Bowers:2001ip,Liu:2002gi,Shovkovy:2003uu,Gubankova:2003uj}.
(The identification of the boundaries
of a blocking region with locations in momentum space
where a dispersion relation is gapless is discussed
with considerable care in Ref.~\cite{Bowers:2001ip}, which considers
a more complicated setting in which rotational symmetry is
spontaneously broken and the blocking regions are not spherically
symmetric.  Such blocking regions were analyzed previously
in Refs. \cite{Alford:2000ze}.)
The pairing criterion \eqn{pairing} can be interpreted as
saying that the free energy cost $2 \De$ of breaking a Cooper
pair of two quarks $a$ and $b$ 
is greater than the free energy $2\de\mu$ gained by
emptying the $a$ state and filling the $b$ state (assuming that
$\de\mu$ pushes the energy of the $a$ quark up and the $b$ quark 
down)~\cite{Rajagopal:2000ff}.  In the blocking region, we
find unpaired $b$ quarks and no $a$ quarks.

We wish now to apply these ideas to the $2\times 2$
pairing blocks in three-flavor quark matter,
first in the CFL phase. As described above,
neutrality is imposed via chemical potentials 
$\mu_e, \mu_3, \mu_8$, and in
the CFL phase the leading effect of the strange quark
mass is an additional effective
chemical potential $-M_s^2/2\mu$ for the strange quarks. The
splittings of the various pairs are then as given in the middle column
of table~\ref{tab:splittings}. 

\begin{table}
\newcommand{\shortstrut}{\rule[-1.5ex]{0em}{4ex}}
\newcommand{\longstrut}{\rule[-2.5ex]{0em}{5.7ex}}
\newlength{\wid}
\settowidth{\wid}{electronless CFL}
\begin{tabular}{ccc}
\hline
\longstrut quark pair & $\de\mu_{\rm eff}$ 
  & \parbox{\wid}{$\de\mu_{\rm eff}$~in\\
     electronless CFL} \\
\hline
\shortstrut $rd$-$gu$& $\frac{1}{2}(\mu_e+\mu_3)$
  & $\mu_e$ \\
\shortstrut $rs$-$bu$ & $\frac{1}{2}(\mu_e+\frac{1}{2}\mu_3+\mu_8 
       - \half M_s^2/\mu)$
  & $\mu_e - \half M_s^2/\mu$ \\
\shortstrut $gs$-$bd$ & $\frac{1}{2}(\frac{1}{2}\mu_3-\mu_8 
       +  \half M_s^2/\mu)$
  & $M_s^2/2\mu$  \\
\hline
\end{tabular}
\caption{
Chemical potential splittings 
for the $2\times 2$ pairing blocks. 
($\de\mu_\mathrm{eff}$ and $\bar \mu$, which is not
tabulated, are defined in each row
such that the effective chemical potentials
of the two quarks that pair are
$\bar \mu \pm \de\mu_\mathrm{eff}$.)
The middle column gives $\de\mu_\mathrm{eff}$ for 
general values of the chemical potentials $\mu_e$, $\mu_3$ and 
$\mu_8$.
In the last column, it is understood that 
as $\mu_e$ is varied, $\mu_3$ and $\mu_8$ ``follow it'' in
such a way that varying $\mu_e$ corresponds to
varying $\mu_{\tilde Q}$, tracking
degenerate $\Qtilde$-neutral solutions for electron-less 
CFL quark matter.
}
\label{tab:splittings}
\end{table}

\begin{figure}[t]

\begin{center}
\includegraphics[width=0.48\textwidth]{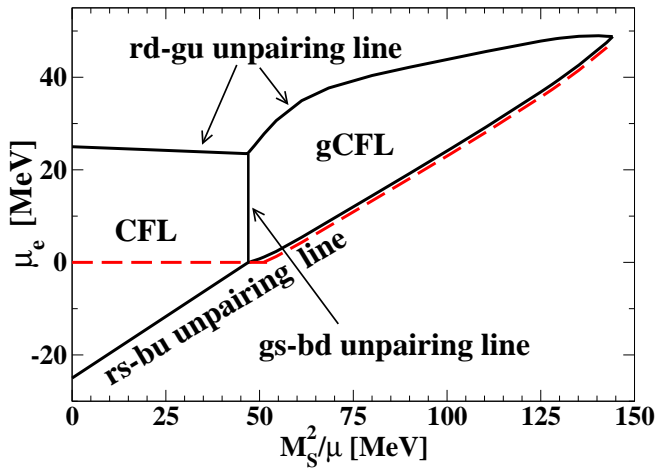}
\end{center}
\vspace{-0.25in}
\caption{Unpairing lines 
for the same parameters as used in Fig.~\ref{fig:gaps}.
If electrons are neglected, then 
the upper and lower curves bound the region of $\mu_e$ 
where neutral solutions to the gap equations are found.
These solutions are all $\tilde Q$-insulators.
Taking electrons into account, the correct solution is the dashed
line: in the CFL phase $\mu_e=0$, while the gCFL phase corresponds
to values of $\mu_e$ below but
very close to the $rs$-$bu$ unpairing line. gCFL 
is a $\tilde Q$-conductor both because
of the nonzero electron density and because of the ungapped $\tilde Q$-charged
$rs$-$bu$ quasiparticles.
}
\label{fig:mues}
\end{figure}

Electrons will play a crucial role in understanding the gCFL
phase, but it is fruitful initially to consider 
matter consisting only of quarks, 
which we can do by sending the electron mass
to infinity. 
In the absence of electrons, at each $M_s^2/\mu$ 
there is a plateau in the free energy of neutral CFL (or gCFL) solutions:
if we vary the chemical potential that couples to $\tilde Q$ charge,
\beq
\mu_{\tilde Q} = -{{\txt \frac{4}{9}}}\left(\mu_e + \mu_3 + \half \mu_8
\right)\ ,
\eeq
while keeping constant the gap parameters
$\De_i$ and the two orthogonal combinations of chemical potentials,
then over a range of  $\mu_{\tilde Q}$ the free energy does
not change and we have a family of neutral stable solutions
to the gap equations.
This indicates that,
in the absence of electrons, both the CFL and gCFL phases are
$\Qtilde$-insulators. 
On this plateau,  all 
$\Qtilde$-charged quasiparticles remain gapped:
\eqn{pairing} is obeyed for the $(rd,gu)$ and $(rs,bu)$
$2\times 2$ quark pairing blocks.
At the edges of the plateau, some $\Qtilde$-charged quasiparticles
become gapless, 
the material ceases to be $\tilde Q$-neutral,
$\partial \Omega /\partial \mu_{\tilde Q}\neq 0$, and the free energy
is no longer independent of changes in $\mu_{\rm \tilde Q}$.
The range of $\mu_\Qtilde$ that defines the plateau is therefore
the band gap for the CFL/gCFL insulator. 
In  Fig.~\ref{fig:mues} we show the unpairing lines for each 
$2\times 2$ quark pairing block. The $rd$-$gu$ line and the
$rs$-$bu$ line bound the plateau region.
Although the vertical axis
is labelled  ``$\mu_e$'', it actually corresponds to variation
in $\mu_\Qtilde$, since we varied
$(\mu_e,\mu_3,\mu_8)$ by a multiple of $(1,1,\frac{1}{2})$.
In the CFL phase, this corresponds to
keeping $\mu_3=\mu_e$ and $\mu_8=\frac{1}{2}(\mu_e-M_s^2/\mu)$
while varying $\mu_e$.

We see that (g)CFL matter exists in a wedge,
between the $rd$-$gu$ unpairing line and the $rs$-$bu$ unpairing
line.
From table \ref{tab:splittings} we can see
that the $bd$-$gs$-unpairing line is vertical because 
the $bd$ and $gs$ quasiparticles are $\tilde Q$-neutral, so
their splitting depends only on $M_s^2/\mu$ and not 
on $\mu_e$. This unpairing line has a different character
than the other two. Rather than bounding the band-gap within
which solutions are found, it separates the CFL and gCFL phases.
CFL is stable only up to a critical value of $M_s^2/\mu$, where the
the $gs$-$bd$ pairs break.

At the lower ($rs$-$bu$-unpairing) line, $\mu_\Qtilde$ is large enough
that the $bu$ and $rs$ quarks, which have $\tilde Q=+1$ 
and $\tilde Q=-1$ respectively and which pair with gap parameter $\De_2$,
no longer pair completely: 
it is energetically
favorable to create a new blocking region of unpaired $bu$ quarks.
At this 
$\tilde Q$-electrostatic potential, the CFL $\tilde Q$-insulator
breaks down, unpaired $bu$ quarks with $\tilde Q=+1$ are created,
the free energy is no longer $\mu_{\tilde Q}$-independent, and
in fact the neutrality conditions and gap equations are no longer
satisfied. 

At the upper ($rd$-$gu$-unpairing) line,
$\mu_{\tilde Q}$ is so low that the $rd$ and $gu$ quarks,
which have $\tilde Q=-1$ 
and $\tilde Q=+1$ respectively and which pair with gap parameter $\De_3$,
no longer pair completely, and
it is energetically
favorable to create a new blocking region of unpaired $rd$ quarks,
and once again no solution is found. 

At $M_s^2/\mu=143$~MeV, which is so large that the gapless CFL
phase is anyway already metastable with respect to unpaired
quark matter, the two boundaries cross, meaning that no gapless CFL
solution can be found. 

So, in the absence of electrons, we can find stable solutions of
the gap and neutrality equations
everywhere between the $rs$-$bu$ and $rd$-$gu$ curves in Fig.~\ref{fig:mues}.
To the left of the $gs$-$bd$ unpairing line this is the CFL phase, 
a $\Qtilde$-insulator with
no gapless quasiquark modes. To the right of that line we have the gCFL phase,
again a $\Qtilde$-insulator, in which all $\Qtilde$-charged
modes are gapped, but there are $\tilde Q=0$
gapless quasiparticles.

We now restore the electrons, 
setting their mass to zero. In the CFL region, the system is forced to 
$\mu_e=0$ (dashed line in Fig.~\ref{fig:mues}) \cite{Rajagopal:2000ff}. 
However,
at the transition point to gCFL,
where the $gs$-$bd$ pairs break, we find that
the neutrality requirement forces us over the line where
$rs$-$bu$ pairs also begin to break.  The result is that 
as $M_s^2/\mu$ increases further, the system
maintains neutrality by staying close to the $rs$-$bu$-unpairing
line, where there is a narrow blocking region in which there
are unpaired $bu$ quarks. Their charge is cancelled by a small
density of electrons. We analyze this quantitatively below.

We see that real-world gCFL quark matter is a conductor of $\tilde Q$
charge, since it has gapless $\tilde Q$-charged quark modes, as
well as electrons. The $rd$ and $gu$ quarks, which are
insensitive to the strange quark mass, remain robustly paired
and the $\tilde Q$-neutral $bd$ and $gs$ quarks develop a large
blocking region as the system moves far beyond their unpairing line.
The neutrality  requirement naturally keeps the system close to
the $rs$-$bu$-unpairing line, following the
dashed line in Fig.~\ref{fig:mues}, so these quarks have a very narrow 
blocking region and an almost quadratic
dispersion relation (see below). 
Although $U(1)_{\tilde Q}$ is unbroken in the gapless CFL phase, the
presence of electrons and unpaired $bu$ quarks makes this phase a
$\tilde Q$-conductor.  
This is in contrast to the CFL phase, which is a $\Qtilde$-insulator
with no gapless quasiquarks and no electrons.

The gapless quark quasiparticles occur in the $gs$-$bd$ and $rs$-$bu$
sectors. Since these will have a dramatic effect on transport
properties, we now discuss them in greater depth.

\subsubsection{The $gs$-$bd$ sector}

In a typical part of the gCFL phase space, 
the 
$\Qtilde$-neutral $gs$-$bd$ sector
is well past its unpairing line, and there is a large
blocking region between momenta $p^{bd}_1$ and $p^{bd}_2$
at which there are gapless excitations, as shown in 
Fig.~\ref{fig:disprel}. 
In the blocking region $p^{bd}_1<p<p^{bd}_2$
there are $bd$ quarks but no $gs$ quarks,
and thus no pairing.
We have confirmed this by direct
evaluation of the difference between the number density
of $bd$ and $gs$ quarks, showing this to be equal to
the volume of the blocking region in momentum 
space. 

Note that even though there is no
pairing in the ground state in the 
blocking region, the dispersion relations are not trivial.
Because the states obtained via the two different 
single particle excitations that are possible (adding
a $gs$ quark or removing a $bd$ quark) mix
via the $\De_1$ condensate, the two dispersion relations
exhibit an ``avoided crossing'' between $p^{bd}_1$
and $p^{bd}_2$. If we neglect the mixing 
among the excitations introduced by $\De_1$,
the gapless excitations just above (below)
$p^{bd}_2$ are $bd$ quarks (holes) and those
just above (below) $p^{bd}_1$ are $gs$ quarks (holes).

It may seem coincidental 
that the value of $M_s^2/\mu$ at 
which the CFL phase becomes gapless is the same as the value
at which $\De_1$ and $\De_2$ separate in Fig.~\ref{fig:gaps}.
Although we do not see a profound reason for this, it is certainly
not a coincidence.  The CFL$\rightarrow$gCFL transition is
triggered by the instability of the CFL phase that occurs when
a $gs$-$bd$ quasiparticle dispersion relation goes gapless,
indicating the instability towards $gs$-$bd$ unpairing and 
the opening up of a blocking region in momentum space,
filled with unpaired $bd$ quarks and with
no $gs$-$bd$ pairing. 
Consequently,
one of the terms in the $\De_1$ gap equation --- that
corresponding to the $gs$-$bd$ block --- is reduced in magnitude because
its integrand vanishes within the
blocking region.  This reduction in the support of the $\De_1$
gap equation integrand causes $\De_1$ to drop.

The ``thickness'' of the $bd$ blocking
region can be considered an order parameter
for gCFL:  for $M_s^2/\mu$ below the critical value
there is no blocking region. Just above the critical value
we can use the results of table \ref{tab:splittings} and \eqn{blocking}
to show that
\beq
p^{bd}_2-p^{bd}_1 \sim 
\De_1^{1/2}\left(\frac{M_s^2}{\mu}-\frac{(M_s^c)^2}{\mu} \right)^{1/2}
\sim (M_s-M_s^c)^{1/2}\ ,
\label{breach}
\eeq
typical behavior for a second order phase transition.  Because
we are analyzing a zero temperature
quantum phase transition, the long wavelength
physics at the
critical point is $4$-dimensional rather than $3$-dimensional 
as at a finite temperature transition.

\subsubsection{The  $rs$-$bu$ sector}

As discussed above, the gCFL phase remains neutral by crossing
the $rs$-$bu$ unpairing line, and developing enough unpaired
$bu$ quarks to cancel the $\Qtilde$ charge of the electrons.
The electrons contribute $(-\mu_e^4/12\pi^2)$ to the free energy,
so $\tilde Q$-neutrality can be maintained as long as
\beq
\label{Qtildeneutrality}
n_e = \frac{\mu_e^3}{3\pi^2} = n_{bu} = \frac{(p^{bu}_2)^3-(p^{bu}_1)^3}
{3\pi^2}\ ,
\eeq
where $p^{bu}_1$ and $p^{bu}_2$ bound the blocking region of unpaired
$bu$ quarks.  The condition (\ref{Qtildeneutrality}) implies that
\beq
\label{bublocking}
(p^{bu}_2 - p^{bu}_1) = \frac{\mu_e^3}{3 \bar p^2}
\eeq
where $\bar p$ is the average of $p^{bu}_1$ and $p^{bu}_2$.
At $M_s^2/\mu=80$~MeV, where $\mu_e=14.6$~MeV at the lower curve
in Fig.~\ref{fig:mues}, this implies $(p^{bu}_2 - p^{bu}_1)=0.0046$~MeV!
Indeed, in Fig.~\ref{fig:disprel} the separation between $p^{bu}_1$
and  $p^{bu}_2$ is invisible, and the dispersion relation
appears to be quadratic about a single gapless point.
To resolve the separation between $p^{bu}_1$
and  $p^{bu}_2$, we did calculations assuming 200 and 500 ``flavors''
of massless electrons.  In these cases,  $(p^{bu}_2 - p^{bu}_1)$
$\sim 1$~MeV and $\sim 3$~MeV, in very good
agreement with the above argument.
Returning to our world with its single electron species,
because  $(p^{bu}_2 - p^{bu}_1)$ is so small, the
value of $\mu_e$ at the true $\tilde Q$-neutral solution
is {\em very} close to that given by the lower curve in Fig.~\ref{fig:mues}.
And, the gaps are {\em very} close to those found in a calculation
done in the absence of electrons. 

From \Eqn{disprel2}, the maximum in the quasiparticle energy
between the two gapless momenta is $E_\mathrm{max}=|\de\mu-\De|$,
so from \eqn{blocking} we can express this in terms of
the width of the blocking region: 
$4(|\de\mu|+\De)E_\mathrm{max}=(p_2-p_1)^2$. For the
$rs$-$bu$ quarks, the blocking region is always very narrow,
so $E_\mathrm{max}\approx (p^{bu}_2-p^{bu}_1)^2/(8 \De_2 )$
which from \eqn{bublocking}
is a small fraction of an electron volt at $M_s^2/\mu=80~\MeV$.
Thus, at any astrophysically relevant temperature, the $rs$-$bu$
dispersion relation can be treated as quadratic about a single
momentum at which it is gapless. 
Indeed, even at $M_s^2/\mu=130$~MeV where the gapless CFL phase ceases
to be the ground state, $\mu_e=40.3$~MeV, 
$(p^{bu}_2 - p^{bu}_1)\sim 0.1$~MeV and the peak of the dispersion
relation between $p^{bu}_1$ and $p^{bu}_2$ is at about 50 eV.
The requirement of $\tilde Q$-neutrality naturally forces
this dispersion relation to be {\em very} close to quadratic,
without requiring fine tuning to a critical point.

\subsubsection{The $rd$-$gu$ sector and the
$3\times 3$ $ru$-$gd$-$bs$ block}

In Fig.~\ref{fig:dispersion_5modes} we show the dispersion relations
for the quasiparticles in the $rd$-$gu$ sector.  One of these becomes
gapless at the upper boundary of the wedge in Fig.~\ref{fig:mues},
but we have seen that in the presence of electrons, the
neutral gapless CFL solution is never near this upper boundary.
Therefore, these dispersion relations are always gapped,
as in the figure.
In Fig.~\ref{fig:dispersion_5modes} we also show the dispersion
relations for the three
quasiparticles from the $3\times 3$ block. These quasiparticles carry
zero $\tilde{Q}$ charge and they always have nonzero gap.
Their smallest gap becomes very small near the rightmost 
tip of the gCFL wedge region
in Fig.~\ref{fig:mues}, but is always greater than 1 MeV in the region
in which gCFL is favored.

\subsection{The gCFL Free Energy Function}

In the previous subsection, we have used the dispersion relations
to delineate the unpairing lines which bound the ranges 
in $\mu_\Qtilde$ where, in the absence
of electrons, $\Qtilde$-insulator solutions are
to be found and which separate the CFL and gCFL phases.
Here, we sketch the behavior of the free energy $\Omega$
in the vicinity of solutions to the gap and neutrality
equations, and see how this behavior changes at 
the unpairing lines.

\begin{figure}[htb]
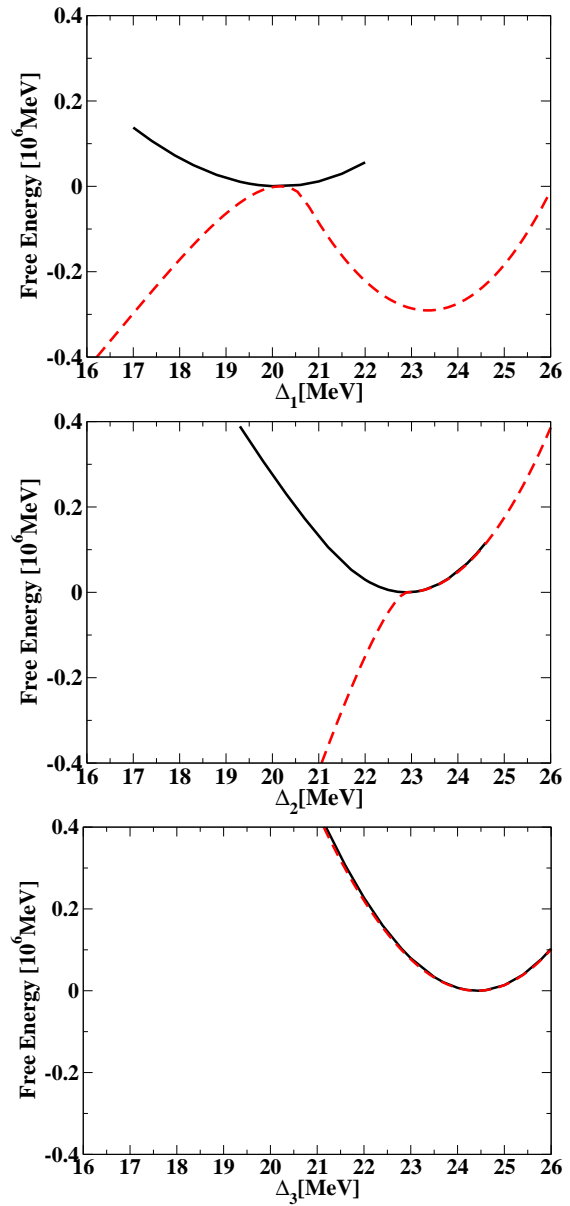

\begin{center}
\includegraphics[width=0.41\textwidth]{de1.eps}
\includegraphics[width=0.41\textwidth]{de2.eps}
\includegraphics[width=0.41\textwidth]{de3.eps}
\end{center}
\vspace{-0.25in}
\caption{These figures show the free energy $\Omega$ in the vicinity
of the gapless CFL solution for $M_s^2/\mu=51.2$~MeV. In each panel,
the dashed curve is obtained by varying one of the gap parameters
($\De_1$ in the top panel, $\De_2$ in the middle; $\De_3$ in
the bottom) while keeping the other two gap parameters and
the chemical potentials $\mu_e$, $\mu_3$ and $\mu_8$ fixed.
The free energies are measured relative to that of the
solution.  The solid curve in each panel depicts the
``neutral free energy'', obtained by varying one gap
parameter, keeping the other gap parameters fixed, and
solving the neutrality conditions anew for each point on
the solid curve.  
}
\label{fig:shovkovyplot}
\end{figure}

\begin{figure}[htb]
\begin{center}
\includegraphics[width=0.41\textwidth]{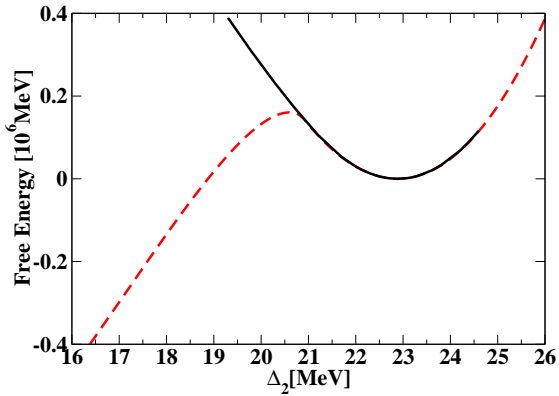}
\end{center}
\vspace{-0.25in}
\caption{
Same as the middle panel of Fig.~\ref{fig:shovkovyplot},
except that $\mu_e$ has been increased by 2 MeV while changing
$\mu_3$ and $\mu_8$ so as to make this a shift in $\mu_\Qtilde$.
This means that, neglecting electrons, we are now exploring
the change of the free energy and the neutral free energy
upon variation of $\De_2$ about
a gCFL solution that is in the interior of
the wedge in Fig~\ref{fig:mues}, rather than at its bottom boundary.
}
\label{fig:shovkovyplot2}
\end{figure}

In Fig.~\ref{fig:shovkovyplot} we study the free energy in
the vicinity of a gapless CFL solution not far above 
the CFL$\rightarrow$gCFL transition.  We have neglected
electrons in making this plot; the change
from including them would be invisible on the
scale of the plot. We plot the free 
energy upon variation of gap parameters while keeping $\mu$'s
fixed (dashed curves in  Fig.~\ref{fig:shovkovyplot}), and 
we also plot the ``neutral free energy'' (solid curves)
obtained by varying gap parameters about the solution while
solving the neutrality conditions anew for each value of the 
gap parameters.
We see that the solution is a minimum
of the neutral free energy, confirming that we have succeeded in finding a 
stable
neutral solution.
However, the solution is not at a minimum
of the free energy upon variation of the gap parameters
while keeping $\mu$'s fixed.   

We see in the top panel of Fig.~\ref{fig:shovkovyplot} that
the solution is found at a local maximum of the dashed curve
describing variation of $\De_1$ at fixed $\mu$'s.
Shovkovy and Huang described similar behavior in
the gapless 2SC case in Refs.~\cite{Shovkovy:2003uu,Huang:2003xd},
and suggested that this is
a characteristic of gapless superconductivity.  We find
this {\em not} to be the case:  deep in the gCFL phase,
at $M_s^2/\mu=80$~MeV rather than the $M_s^2/\mu=51.2$~MeV of
Fig.~\ref{fig:shovkovyplot}, we find that the gCFL solution
is a local minimum of both the dashed and the solid curves
in the analogue of the top panel of Fig.~\ref{fig:shovkovyplot}.
That is, we find an onset of
the behavior seen in the 
top panel of Fig.~\ref{fig:shovkovyplot} 
as we cross the CFL$\rightarrow$gCFL
transition, namely the $gs$-$bd$ unpairing line:
as a $gs$-$bd$ blocking region begins to open up,
the solution
goes from being a local minimum of the dashed curve
to being a local maximum. 
However, we find that the dashed curve 
does not persist in this shape as the 
$gs$-$bd$ blocking region
expands.  The {\em onset} of gaplessness is
characterized by
a dashed curve as in the top panel of
Fig.~\ref{fig:shovkovyplot}, but gaplessness 
itself need not be. 

In the middle panel of Fig.~\ref{fig:shovkovyplot}, 
we find that the solution is at
a point of inflection with respect to variation of 
$\De_2$ at fixed $\mu$'s. We find that
the gCFL solutions 
at all values of $M_s^2/\mu$ above $(M_s^2/\mu)_c$
are at points of inflection of this sort.
This arises
because a gCFL solution is forced
by the neutrality constraint to be very close to
the $bu$-$rs$ unpairing line. In Fig.~\ref{fig:shovkovyplot2}
we replot the 
middle panel of  Fig.~\ref{fig:shovkovyplot} after 
increasing $\mu_e$ by 2 MeV while
varying $\mu_3$ and $\mu_8$ so that only  $\mu_\Qtilde$ changes.
This means that we have taken a 2 MeV step
upwards in Fig.~\ref{fig:mues}, 
away from the $bu$-$rs$ unpairing line. And, we see
that the solution is now a minimum with respect to
variation of $\De_2$ at fixed $\mu$'s.  The point
of inflection has resolved itself into a minimum
and a maximum, with the solution at the minimum. 
Thus, the point of inflection in the dashed curve 
does indeed occur at the $bu$-$rs$ unpairing line.

Note that at $M_s^2/\mu=80$~MeV,
once we have taken an upward step away from the $bu$-$rs$ unpairing line
in Fig.~\ref{fig:mues}, obtaining the analogue of 
Fig.~\ref{fig:shovkovyplot2},
the gapless CFL phase solution 
is now a local minimum of both the dashed and solid curves for
variation in the $\De_1$, $\De_2$ and $\De_3$ directions.

If we take a step in the ``wrong direction'' in Fig.~\ref{fig:mues},
downwards from the $bu$-$rs$ unpairing line, the 
point of inflection in the middle panel of Fig.~\ref{fig:shovkovyplot}
vanishes and the dashed curve becomes
monotonically increasing, indicating that there is no
solution to the $\De_2$ gap equation to be found at 
these values of the $\mu$'s.  
In the presence of electrons,
the neutrality conditions are satisfied {\em just} below the 
$bu$-$rs$ unpairing line in Fig.~\ref{fig:mues}, and the 
dependence of the free energy on $\De_2$ is 
slightly modified so that the point of inflection in 
the middle panel of 
Fig.~\ref{fig:shovkovyplot} occurs where the neutrality
conditions are satisfied.  We have confirmed this in 
calculations done with 200 and 500 species of electrons;
with a single species as in the real world, the changes
in Fig.~\ref{fig:shovkovyplot} are invisible on the scales
of the plot.

Finally, with respect to variation of $\De_3$, the solution is a local 
minimum of the dashed curve in the lower panel
of Fig.~\ref{fig:shovkovyplot}. However, we have verified that
if we move sufficiently upwards in Fig.~\ref{fig:mues} as to run into
the $rd$-$gu$ unpairing line, then
the dashed curve in the lower panel
exhibits a point of inflection (while that in the middle
panel has a robust minimum.)

\subsection{Mixed Phase Alternatives}

\begin{figure}[t]
\begin{center}
\includegraphics[width=0.475\textwidth]{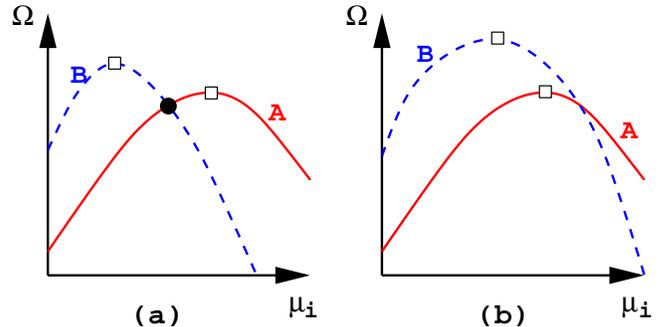}
\end{center}
\vspace{-0.25in}
\caption{
Schematic illustration of conditions for the
occurrence of mixed phases. Free energy  $\Om$ 
for two phases $A$ and $B$ is
shown as a function
of some chemical potential $\mu_i$. Charge $Q_i=-\p\Om/\p\mu_i$
is given by the slope. Squares mark the neutral points.
Panel (a): at the neutral point for each phase,
the other phase has lower free energy,
so there is a point (black dot) where the two phases can coexist with the same 
pressure and opposite charge, with lower free energy than either neutral phase.
Depending on Coulomb and
surface energy costs, a mixed phase may exist there.
Panel (b): phase $B$ has higher free energy
than phase $A$ at the point where $A$ is neutral. At no point
do the two phases coexist with opposite charge, so no mixed phase
is possible.
}
\label{fig:mixed}
\end{figure}

Up to this point, the phases we have discussed have been
locally neutral with respect to all gauge charges. However,
it is well known that neutrality can also be achieved in
an averaged sense,
by charge separation into a mixture of two oppositely charged
phases. This is shown schematically in Fig.~\ref{fig:mixed},
which shows generic free energy curves $\Om(\mu_i)$ for two phases $A$ and $B$.
The free energy must be a concave function of the chemical potential,
since increasing $\mu_i$ increases the charge $Q_i=-\p\Om/\p\mu_i$.
There are then two possible situations. In one (Fig.~\ref{fig:mixed}b)
there is no  coexistence point and hence no mixed phase is possible.
In the other (Fig.~\ref{fig:mixed}a) 
there is a coexistence point of oppositely-charged
phases, and its free energy is lower than that 
of either neutral phase, so if Coulomb
and surface energy costs
are low enough then a neutral mixed phase will be free-energetically preferred
over either homogeneous neutral phase.

We now consider possible gCFL+unpaired mixed phases.
(Note that for $(M_s^2/\mu)>(M_s^2/\mu)_c$
a CFL+unpaired mixture is not possible because
there is no CFL solution, charged or neutral.)
For the unpaired and gCFL phases, the free energies are of the form
shown in Fig.~\ref{fig:mixed}a.
At the values
of $(\mu_e,\mu_3,\mu_8)$ that make one phase neutral, the other phase has
lower free energy. Thus there is a 
value of $(\mu_e,\mu_3,\mu_8)$ ``between'' that for
neutral gCFL and that for neutral unpaired quark matter,
where the two can coexist with opposite color and electric charge density.
However, a mixed phase is not favored in this case
because each component would have net color charge, and 
color is a gauge symmetry with a strong coupling
constant, so this
mixed phase would pay a huge 
price in color-Coulomb energy.
(For similar arguments applied to systems with no gauge symmetries,
where the initial conclusion that a mixed phase is
favored is the correct one, 
see Refs.~\cite{Bedaque:2003hi,Forbes:2004cr}.)

It is then natural to ask whether one could
construct a gCFL+unpaired mixed phase whose components
are electrically charged, but color neutral. This avoids the
large color-Coulomb energy cost of mixed phases
with colored components. Such mixed phases
have recently been constructed in two-flavor quark matter, with
unpaired and 2SC components \cite{Reddy:2004my}.
There is still some color electric field (and
color-charged boundary layers) at the interfaces
where the color chemical potentials change rapidly as one travels
from one component to the other, analogous to the
charged boundary layers and ordinary electric field
at the CFL/nuclear interface constructed in Ref.~\cite{Alford:2001zr},
but Ref.~\cite{Reddy:2004my}
finds that the 2SC + unpaired mixed phase does occur in two-flavor
quark matter. 
However, for color neutral unpaired and gCFL phases,
we have found that the situation is typically that of Fig.~\ref{fig:mixed}(b):
the free energy of color neutral, but electrically charged,
unpaired quark matter is typically {\em higher} than
the free energy of color-neutral gCFL, at the value of $\mu_e$
where color-neutral gCFL is electrically neutral.
Hence there is no value of $\mu_e$ at which
oppositely charged phases can coexist.
We have found that this is true for all values of $M_s^2/\mu$
except for a range of a few MeV just below $M_s^2/\mu=130$~MeV,
where the neutral gCFL and neutral unpaired free energies cross.
There, a mixed phase may arise, although as we shall discuss below
it may be superseded by other more favorable possibilities.

We have not eliminated all possible mixed phase constructions,
involving mixtures of all possible phases.  
However, over most of the gCFL regime there can be no mixed phase
constructed from gCFL and unpaired quark matter.

\subsection{(Gapless) 2SC and 2SCus}
\label{sec:g2SC}

\begin{figure}[t]
\begin{center}
\includegraphics[width=0.48\textwidth]{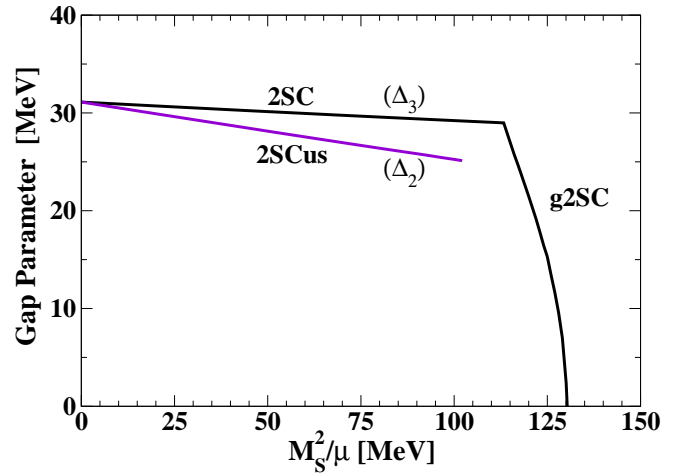}
\end{center}
\vspace{-0.25in}
\caption{Gap parameters in 3-flavor quark matter
for the 2SC phase $(\De_3)$ and the
2SCus phase ($\De_2$). In each case the other two gaps are zero.
The 2SC phase becomes gapless (g2SC) at $M_s^2/\mu>113$~MeV
and ceases to exist at $M_s^2/\mu\approx 130$~MeV, and its free energy
is always lower than that of unpaired quark matter (Fig.~\ref{fig:energy}).
The 2SCus phase becomes unfavored relative to unpaired quark matter
at $M_s^2/\mu>99$~MeV, and ceases to exist at $M_s^2/\mu\approx 103$~MeV,
without ever becoming gapless.
}
\label{fig:gapless2sc}
\end{figure}

In this subsection, we discuss the properties of phases in which
only two of the three flavors pair. These cannot compete with the CFL and
gCFL phases at low values of $M_s^2/\mu$, but could conceivably become 
important at larger values (lower densities).

The Fermi momenta in cold unpaired
quark matter are ordered $p_{Fd} > p_{Fu} > p_{Fs}$, since the
strange quark mass tends to decrease the strange quark Fermi momentum, 
and the down quark Fermi momentum then increases to preserve neutrality.
Thus, the likely two-flavor pairings in cold three-flavor quark matter
are $u$-$d$ pairing (i.e. 2SC, with gap parameter $\De_3>0$ and
$\De_1=\De_2=0$) and  $u$-$s$ pairing (i.e.~2SCus, with gap parameter
$\De_2>0$ and $\De_1=\De_3=0$).

\subsubsection{Calculation}
In order to find a two-flavor pairing solution, we need only solve
four equations (one gap and the three neutrality equations). The other
two gap equations are automatically satisfied upon setting
the relevant gaps to zero. Using the same coupling strength as in our
investigation of the gCFL phase ($\De_0=25~\MeV$) and working at
the same value of $\mu=500~\MeV$, 
the nonzero gaps at $M_s^2/\mu=0$ are $\De_3=31$~MeV in the 
2SC phase and $\De_2=31$~MeV in the 2SCus phase.
As we increase $M_s^2/\mu$,
as long as we do not enter a gapless phase the gaps
decrease slowly
and the 
simplified analysis of the 2SC and 2SCus phases in
Ref.~\cite{Alford:2002kj} should
be a good guide. We do indeed find that 
our results are well approximated by 
$\mu_3=\mu_8= 0$ and $\mu_e= M_s^2/2\mu$ in the
2SC phase and $\mu_e=\mu_3=\mu_8= 0$ in the 2SCus
phase, with a free energy given by
\begin{eqnarray}
\Omega_{\mbox{\scriptsize\begin{tabular}{l}neutral\\ 
2SC/2SCus
\end{tabular}}}
&=& \Omega_{\mbox{\scriptsize\begin{tabular}{l}neutral\\ unpaired\end{tabular}}}+ \frac{ M_s^4 
- 16 \Delta_i^2\mu^2}{16\pi^2}\ ,
\label{2SCresult}
\end{eqnarray}
with $\De_i$ given by 
$\De_3$ in the 2SC phase and $\De_2$ in the 2SCus phase,
as predicted in Ref.~\cite{Alford:2002kj}.  The free energy
of the 2SCus phase is higher than that of the 2SC phase
because $\De_2$ decreases more
rapidly with $M_s^2/\mu$ in the 2SCus phase than $\De_3$
does in the 2SC phase. This cannot be discovered by
the methods of Ref.~\cite{Alford:2002kj}, in which
these two phases were treated as degenerate.  

Our results for the gap parameters are shown in Fig.~\ref{fig:gapless2sc}
and for the free energies in Fig.~\ref{fig:energy}. As in our other figures,
we vary $M_s$ keeping $\mu$ fixed at $500$~MeV.

\subsubsection{2SC/g2SC Results}

We find a neutral 2SC solution at low $M_s^2/\mu$, with
four gapped quasiparticles. At $M_s^2/\mu\approx 113$~MeV
two of these quasiparticles become gapless,
with blocking regions within which there are
unpaired $rd$ and $gd$ quarks, 
and there is a
continuous transition to the gapless 2SC (g2SC) phase.
(Gapless 2SC was introduced in      
two-flavor quark matter in Refs.~\cite{Shovkovy:2003uu,Huang:2003xd}.)
The gap parameter then decreases rapidly until it reaches zero
at $M_s^2/\mu\approx 130$~MeV and the solution ceases to exist.

As is clear from Fig.~\ref{fig:energy},
2SC/g2SC always has lower free energy than unpaired quark matter,
and usually has higher free energy than CFL/gCFL.
However, we find a tiny
window of $M_s^2/\mu$ less than 1 MeV wide,
very close to $130$ MeV, in which
the gapless 2SC phase has lower free energy than gCFL.
In this regime, the one nonzero gap in the g2SC phase
is almost zero whereas all three gaps
are nonzero in the gCFL phase. This indicates that 
the fact that the gCFL free energy crosses that
of unpaired quark matter almost at the same point
where the g2SC and unpaired free energies come together
is a nongeneric feature of our model.
Taken literally, our calculation predicts
that as $M_s^2/\mu$ increases, gCFL is supplanted
by g2SC which is then almost immediately supplanted by
unpaired quark matter.  However,
treating the effects of $M_s$ more accurately than
we have may shut the tiny g2SC window 
completely~\cite{kenji:private}.  In contrast, treating
$M_s$ as a chemical potential shift, as we have, but using 
$\De_0=100$~MeV appears to open a wide g2SC window~\cite{Ruester:2004eg},
but this occurs in a regime where $M_s\sim\mu$ and so this
result is not trustworthy.  Also, as we discuss in the
next section, a more general ansatz is required once
one is at a sufficiently large $M_s^2/\mu$ that the
free energy of the gCFL phase is close to that of unpaired
quark matter, since there are other possible pairing patterns
that likely become favorable.

Finally, it is interesting to note that
the 2SC solution in three-flavor quark matter differs from
its two-flavor version, which requires a large $\mu_e$
for neutrality given that there are no strange quarks
present to carry negative charge.  In three-flavor 
quark matter, we find that
at $M_s=0$ there is a small positive $\mu_e$ and a small
negative $\mu_8=-\mu_e$ in the 2SC phase. This happens because the pairing of 
$ru$, $rd$, $gu$ and $gd$ quarks increases their number density.
This contributes a positive electric charge and excess redness/greenness,
which is compensated by a small positive $\mu_e$
and a negative $\mu_8$.  As we increase $M_s^2/\mu$,
the small $\mu_8$ remains approximately unaffected whereas the small $\mu_e$
due to pairing is rapidly swamped by the larger contribution
of order $M_s^2/2\mu$ that compensates for the lack of strange quarks.

\subsubsection{2SCus results}

We find a neutral 2SCus solution, with a gap $\De_2$ that decreases
with increasing $M_s^2/\mu$ as shown in Fig.~\ref{fig:gapless2sc}.  
This solution only exists for $M_s^2/\mu<103$~MeV, and
has a higher free energy than that of neutral unpaired quark
matter for $M_s^2/\mu>99$~MeV (see Fig.~\ref{fig:energy}).
It is always unfavored relative to the CFL/gCFL phase.

It is striking that two-flavor $u$-$s$ pairing,
unlike the two-flavor $u$-$d$ pairing discussed above, has no gapless
phase. In our calculations, we find that at $M_s^2/\mu>103$~MeV,
when the 2SCus phase
becomes unstable to unpairing (i.e. when $\de\mu_{\rm eff}$ in the $u$-$s$
sector, line 2 of table \ref{tab:splittings}, becomes as large as $\De_2$), 
there is no neutral solution with a smaller value
of $\De_2$ (other than unpaired quark matter).
We do find such a ``gapless 2SCus'' solution
in a range of $M_s^2/\mu$ below 103 MeV, 
with $\De_2$ smaller than that in the 2SC solution at the
same $M_s^2/\mu$,
but the g2SCus solution is unstable: it is a local maximum of the
neutral free energy as a function of $\De_2$.
(In a figure like
Fig.~\ref{fig:shovkovyplot} this g2SCus solution 
would be at a local maximum of the {\em solid} curve.)
At $M_s^2/\mu=103$~MeV,
the g2SCus solution (local maximum) meets the 
2SCus solution (local minimum) at an inflection point
of the neutral free energy, and for $M_s^2/\mu>103$~MeV
neither 2SCus nor g2SCus solutions exist.
Unlike the gapless 2SC phase \cite{Shovkovy:2003uu,Huang:2003xd}
and the gapless CFL phase \cite{Alford:2003fq}, which are rendered stable
by the constraint of neutrality, the gapless 2SCus phase
remains unstable. This is presumably because $u$-$s$ paired
phases are very close to being neutral anyway (only very small
values of $\mu_e,\mu_3,\mu_8$ are required to achieve neutrality
\cite{Alford:2002kj}),
so the constraint does not change the physics much.
This can be summarized by saying that the 2SCus phase
behaves analogously to that studied by Sarma
\cite{Sarma}, even after neutrality constraints
are imposed.

\section{Concluding Remarks and Open Questions}

The gapless CFL phase seems sufficiently well-motivated
as a possible component of compact stars to
warrant further study of its low energy properties and its 
phenomenological consequences: it is
the phase that supplants the asymptotic CFL phase
as a function of decreasing density, and compact stars
are certainly far from asymptotically dense.

The low energy effective theory of the gCFL phase
must incorporate the gapless fermionic
quasiparticles with quadratic dispersion relations,
which have number densities $\sim \mu^2 \sqrt{\De_2 T}$
and dominate the low temperature specific heat,
the gapless quarks with linear dispersion relations,
with number densities $\sim\mu^2 T$, and the electron
excitations, with 
number density $\sim \mu_e^2 T$. In contrast, the (pseudo-)Goldstone
bosons present in both the CFL and gCFL phases have
number densities at most $\sim T^3$.
This means the gCFL phase will have very different phenomenology
from CFL.
It will be particularly interesting
to compute the cooling of a compact star with a gCFL
core, because neutrino emission will require conversion
between quasiparticles with linear and quadratic 
dispersion relations. And, we expect that in a star with
CFL, gCFL and nuclear volume fractions, the gCFL shell
will dominate the total heat capacity and the total
neutrino emissivity, and thus control the (rapid) cooling.
It will also be interesting to work out the
magnetic field response of the gCFL phase, since the
gauge boson propagators will be affected both
by the gapless quasiparticles (all nine gauge bosons)
and by the condensate (Meissner effect for eight out of nine.)
Finally, we have left the study of possible meson condensation
in the gapless CFL phase to future work.

Although we have studied
the gCFL phase in a model, all of the qualitative
properties of this phase 
that we have focussed on appear robust. We
have also offered a model-independent argument for the instability
that causes the transition, and for the location of
the transition.  We have used our model to show that the gCFL
phase is favored over the
two-flavor-pairing phases (2SC, g2SC, and 2SCus)
throughout almost
all of the regime where the gCFL phase is favored over
unpaired quark matter.
It remains a possibility, however,
that the CFL gap is large enough that 
baryonic matter
supplants the CFL phase before $M_s^2/\mu>2\De$.
Assuming that the gCFL phase does replace the
CFL phase, it is also possible that gaps are small enough that
a third phase of quark matter could supplant the gCFL
phase at still lower density, 
before the transition to baryonic matter.
We do not trust our analysis to determine this third 
phase.  Perhaps it is a mixed phase of some sort,
although we have ruled out the straightforward
possibilities.  Perhaps it is the gapless 2SC 
phase~\cite{Shovkovy:2003uu,Huang:2003xd}, as the
literal application of our model would suggest.
We should stress, in addition, that our model relies upon
a pairing ansatz designed to study the instability of
the CFL phase, and hence well-suited to the study of
the gCFL phase.  Determining what phase comes after
gCFL almost certainly requires a more general ansatz.
For example, perhaps weak pairing
between quarks with the same flavor plays a role
once gCFL is superseded~\cite{Alford:2002rz},
or perhaps it is the crystalline color superconducting
phase~\cite{Alford:2000ze,Kundu:2001tt,Bowers:2001ip,Bowers:2002xr,Casalbuoni:2004wm} 
that takes over from gapless CFL at lower
densities.  (Other possibilities have also been 
suggested~\cite{Muther:2002ej}.)

Recent developments~\cite{Casalbuoni:2004wm} make the 
crystalline color superconducting phase look like
the most viable contender for the ``third-from-densest phase''.
Previous work~\cite{Bowers:2002xr} had suggested that the
face-centered-cubic crystal structure was sufficiently 
favorable that its free energy could be competitive with
that of BCS pairing over a wide range of parameter space,
but because these indications came from a Ginzburg-Landau
calculation pushed beyond its regime of validity, quantitative
results were not possible. The 
results of Ref.~\cite{Casalbuoni:2004wm} suggest that a
crystalline phase involving pairing of only two flavors
is favored over the unpaired phase by 
$\approx 0.2 \mu^2\De_{\rm 2SC}^2/\pi^2$ at 
$M_s^2/\mu \approx 4 \De_{\rm 2SC}$. Here,
$\De_{\rm 2SC}$ is the gap parameter in the 2SC phase
at $M_s=0$, which is $31$ MeV with the parameter values
we have used in all our figures.  This suggests that
if we were to generalize our pairing ansatz to
allow the crystalline phase as a possibility, it
would take over from gCFL at $M_s^2/\mu\sim 120$~MeV,
or even somewhat lower if the three-flavor crystalline
phase, which no one has yet constructed, is more favorable
than the two-flavor version.  Furthermore, the authors of
Ref.~\cite{Casalbuoni:2004wm} find that the crystalline
phase persists until
a first order crystalline$\rightarrow$unpaired transition
at $M_s^2/\mu \approx 7.5 \De_{\rm 2SC}$, hence over a very wide
range of densities.  
If analysis of  three-flavor crystalline color superconductivity
supports these estimates, we will not have to worry about
the resolution of the puzzles and possible
mixed phases associated with the confluence of
the free energies for the gCFL, g2SC
and unpaired phases near $M_s^2/\mu\sim130$ MeV in
Fig.~\ref{fig:energy}: by that density the crystalline
phase will already be robustly ensconced on the phase
diagram.

\begin{acknowledgments}
We acknowledge helpful conversations with J. Bowers, R. Casalbuoni,
G. Cowan, M. Forbes, K. Fukushima, E. Gubankova, J. Kundu,
W. V. Liu, G. Nardulli, S. Reddy, T. Sch\"afer, I. Shovkovy and F. Wilczek,
and are grateful to the INT at the University of
Washington in Seattle for its hospitality.
This research was supported in part by DOE grants 
DE-FG02-91ER40628 and DF-FC02-94ER40818.
\end{acknowledgments}

\end{document}